\documentclass[aps,prl,reprint,superscriptaddress]{revtex4-2}
\usepackage[utf8]{inputenc}

\usepackage[hidelinks]{hyperref}

\usepackage{hyperref}
\usepackage{color}
\hypersetup{colorlinks=true}
\usepackage{graphicx}
\usepackage{bm}
\usepackage{amsmath}
\usepackage{amssymb}
\usepackage{xspace}
\usepackage[capitalise]{cleveref}
\usepackage{txfonts}

\usepackage{physics}
\usepackage{siunitx}
\usepackage{chemformula}
\newcommand{\subfigref}[2]{Fig.~\hyperref[#1]{\ref*{#1}#2}}

\newcommand{\jwone}[1]{j^{\sigma}_{#1(\omega)}}
\newcommand{\jwtwo}[1]{j^{\sigma}_{#1(2\omega)}}
\newcommand{\jtwo}[1]{j^{\sigma}_{#1(0)}}

\newcommand{\PZT}{\mathrm{PZ}}
\newcommand{\strain}{\mathrm{strain}}

\newcommand{\alphae}{0.4}

\newcommand{\density}{\SI{1.5e+28}{m^{-3}}}

\newcommand{\sigmae}{\SI{e+06}{S/m}}

\newcommand{\Ezero}{\SI{58}{kV/m}}

\newcommand{\utophizero}{\SI{14}{V/nm}}

\newcommand{\thickness}{\SI{10}{nm}}
\newcommand{\SAWspeed}{\SI{5}{km/s}}

\newcommand{\SAWomega}{\SI{0.5}{GHz}}
\newcommand{\taue}{\SI{2}{fs}}
\newcommand{\meff}{0.4}

\newcommand{\alphag}{\num{e-03}}
\newcommand{\alpham}{\SI{1.2}{meV.nm^2}}

\newcommand{\taum}{\SI{13}{ps}}
\newcommand{\piezovarrho}{0.01}
\newcommand{\sigmam}{\SI{e+05}{S/m}}
\newcommand{\vsmagnon}{\SI{45}{meV.\AA}}
\newcommand{\thicknessplatinum}{\SI{5}{nm}}
\newcommand{\hallangle}{\num{0.1}}
\newcommand{\leadlength}{\SI{500}{\micro m}}
\newcommand{\rhoplatinum}{\SI{50}{\micro\ohm \cm}}
\newcommand{\htheta}{0.217+0.183\varrho^2+0.274\cos4\theta + 0.0574 \cos8\theta}
\newcommand{\zetaelectron}{\num{e+04}}
\newcommand{\Izeropropto}{\SI{6}{mW/(m GHz)}}

\newcommand{\chimagnon}{\num{e-05}}
\newcommand{\deformation}{\num{e-06}}
\begin{document}
	
	\title{Surface acoustic wave driven acoustic spin splitter in $d$-wave altermagnetic thin films}
	
	\author{Pieter M. Gunnink}
	\email{pgunnink@uni-mainz.de}
	
	\author{Jairo Sinova}
	
	\author{Alexander Mook}
	
	\affiliation{Institute of Physics, Johannes Gutenberg-University Mainz, Staudingerweg 7, Mainz 55128, Germany}
	\date{\today}
	\begin{abstract}
		The generation of spin currents is a key challenge in the field of spintronics. We propose using surface acoustic waves (SAWs) to generate spin currents in altermagnetic thin films, thereby realizing an acoustic spin splitter. Altermagnets, characterized by spin-polarized electrons and magnons, provide a versatile platform where SAWs can drive spin currents carried by both charge carriers and magnons. This acoustic spin splitter can be implemented in both metallic and insulating altermagnetic thin films, offering broad material applicability and a novel way to detect the spin splitter effect in insulating altermagnetic thin film. We examine a realistic experimental setup where a heavy metal layer, such as platinum, is used to convert the spin current into a measurable charge current via the inverse spin Hall effect. For representative material parameters, we calculate the expected spin current and the corresponding inverse spin Hall voltage. Furthermore, we demonstrate that tuning the SAW frequency allows for precise control over the spin current, highlighting the versatility and potential of the acoustic spin splitter for future spintronics applications.
	\end{abstract}
	\maketitle

	\paragraph{Introduction.} Spin currents, the flow of angular momentum, are a cornerstone of spintronics. Traditionally, generating spin currents has relied on either strong spin-orbit coupling (SOC) via the spin Hall effect \cite{sinovaSpinHallEffects2015} or ferromagnetic materials, which present challenges in terms of miniaturization \cite{manchonCurrentinducedSpinorbitTorques2019}.
	Altermagnets offer an alternative approach to generate spin currents in the nonrelativistic limit, making use of collinear and compensated magnetic order \cite{ smejkalConventionalFerromagnetismAntiferromagnetism2022, smejkalCrystalTimereversalSymmetry2020, smejkalEmergingResearchLandscape2022}. Previously, the altermagnetic generation of spin current has been considered in the context of organic salts \cite{nakaSpinCurrentGeneration2019} and metallic oxides \cite{gonzalez-hernandezEfficientElectricalSpin2021,sourounisEfficientGenerationSpin2024}. 
	
\begin{figure}
		\centering
		\includegraphics[width=\columnwidth]{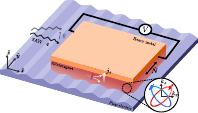}
		\caption{A surface acoustic wave (SAW) is excited in a piezoelectric at an angle $\theta$ with an altermagnetic thin film. The SAW generates a deformation and piezoelectric field, driving a transverse spin current $J_s$ in the altermagnetic thin film and into the heavy metal layer, where it generates a voltage through the inverse spin Hall effect. The inset shows the altermagnetic spin splitting of the Fermi surfaces.} \label{fig:setup}
	\end{figure}

	Herein, we propose to make use of surface acoustic waves (SAWs) to generate spin current. 
	SAWs are a versatile tool to study transport phenomena, such as the acousto-electric effect \cite{parmenterAcoustoElectricEffect1953, weinreichObservationAcoustoelectricEffect1957,kalameitsevValleyAcoustoelectricEffect2019,sonowalAcoustoelectricEffectTwodimensional2020,bhallaPseudogaugeFieldDriven2022} and single electron transport \cite{talyanskiiSingleelectronTransportOnedimensional1997}. In addition, SAWs can also interact with the excitations of the magnetic order, transporting angular momentum \cite{uchidaLongrangeSpinSeebeck2011,uchidaAcousticSpinPumping2012,adachiTheorySpinSeebeck2013a,uchidaSurfaceacousticwavedrivenSpinPumping2011,sanoAcoustomagnonicSpinHall2024}. 
	We consider an altermagnetic thin film, on top of a piezoelectric substrate. By exciting a SAW in the piezoelectric, the spin splitting effect in the altermagnet can be driven acoustically, resulting in a  spin current flowing along the film normal---as sketched in Fig.~\ref{fig:setup}.
	
	There are two main carriers of spin in altermagnets: electrons and magnons, the excitations of the magnetic order. Both exhibit the altermagnetic band splitting \cite{nakaSpinCurrentGeneration2019,gonzalez-hernandezEfficientElectricalSpin2021,smejkalEmergingResearchLandscape2022,smejkalConventionalFerromagnetismAntiferromagnetism2022,smejkalChiralMagnonsAltermagnetic2023} and can facilitate a non-relativistic spin splitter to generate spin current.
	We consider both metallic thin films, where electrons are the main carriers of spin, and insulating thin films, where magnons are the main carriers of spin, and find both quasiparticles can be acoustically driven to generate a transverse spin splitter---making it a versatile tool to study the altermagnetic spin splitting. In particular, insulating altermagnets might offer a clean platform to directly probe the altermagnetic spin splitting, since residual electric signals would not be present.

	\textit{Setup. }
	We assume a SAW excited in a piezoelectric, on top of which an altermagnetic thin film is deposited. The deformation of the SAW then generates a piezoelectric field, which extends into the altermagnetic thin film and couples to the electrons' charge. In addition, the deformation itself can extend into the thin film, coupling to both electrons and magnons. However, the deformation force acting on the electrons is typically much weaker than the piezoelectric field, and thus does not give a leading contribution to the electron dynamics. The deformation is however the sole driving force for the magnons, since magnons are electrically neutral. 
	
	We consider an altermagnetic thin film where the N\'{e}el vector is in the plane and perpendicular to the SAW propagation direction, as has recently been achieved in \ch{RuO2} thin films \cite{liaoSeparationInverseAltermagnetic2024}. The spin-split altermagnetic band structure will then support a spin current along the film normal, polarized along the direction of the N\'{e}el vector. 
	To detect the spin current, we propose to use a thin heavy metal layer on top of the altermagnetic film, which will generate a longitudinal charge current through the inverse spin Hall effect \cite{sinovaSpinHallEffects2015,kawadaAcousticSpinHall2021}.

	Different types of SAWs can be excited, which have different non-zero components of the stress tensor $u_{\alpha\beta}$, and which either generate piezoelectric fields or not---depending on the substrate used. To illustrate the versatility of SAWs, we consider here a Bluestein-Gulyaev (BG) SAW \cite{bleusteinNewSurfaceWave1968,gulyaevElectroacousticSurfaceWaves1969}, a piezoelectric variation of a shear horizontal wave, excited with frequency $\omega$ and propagating with a wave vector $\bm q$ in the plane. A BG SAW has only non-zero displacement along $y$, given by $u_y=u_0 e^{-|z|/\lambda_q}e^{i(qx-\omega t)}$, generating a piezoelectric potential $\phi = c_{\mathrm{PZ}}u_0 e^{-|z|q}e^{i(qx-\omega t)}$. Here $u_0$ is the displacement amplitude, $c_{\mathrm{PZ}}$ is a proportionality constant and $\lambda_q$ is the deformation decay length. We give further details in Sec.~VII in the SM \cite{Note1}.
	This SAW thus generates an in-plane piezoelectric field $E=-\partial_x \phi$ along $x$ and the only non-zero components of the stress tensor are $u_{xy}$ and $u_{yz}$. This allows a BG SAW to acoustically induce magnonic (through the lattice deformation) and electronic (through the piezoelectric field) transverse spin currents. We note however that the mechanism considered here is general, and can also be implemented with other types of SAWs, such as Rayleigh waves \cite{dreherSurfaceAcousticWave2012}, which we also expect to be able to drive spin splitter effects.

	\textit{Formalism. } 
	We assume a $d$-wave altermagnet, oriented such that its altermagnetic splitting is in the $xz$-plane and write its low-energy electron dispersion as
	\begin{equation}
		\hbar\omega_{\bm k,e}^\sigma=\frac{\hbar^2 k^2}{2m^*} + \sigma\alpha_e k_xk_z,
	\end{equation}
	where $m^*$ is the effective electron mass and $\alpha$ is the altermagnetic band splitting parameter. The magnon low-energy dispersion follows a similar structure, 
	\begin{equation}
		\hbar\omega_{\bm k,m}^\sigma=v_s k + \sigma\alpha_m k_xk_z,
	\end{equation}
	but the magnon dispersion is linear in $k$. Here $v_s$ is the magnon velocity and $\alpha_m$ is the magnonic altermagnetic band splitting parameter. For both magnons and electrons, the spin splitting is maximal along the $|k_x|=|k_z|$ diagonal and vanishes along $k_x=0$ or $k_z=0$. 
	
	We now  discuss the dynamics of the electrons and magnons, described by the Boltzmann equation \cite{marderCondensedMatterPhysics2010} 
	\begin{equation}
		\partial_t f_\eta^\sigma + \bm v_{\eta\bm k}^\sigma \cdot \bm\nabla_{\bm r}f_\eta^\sigma + \bm F_\eta\cdot \bm\nabla_{\bm k} f_\eta^\sigma = -\frac{1}{\tau_\eta} \left(f_\eta^\sigma - \langle f^\sigma_\eta \rangle\right), 
	\end{equation}
	where $f_\eta^\sigma$ is the distribution of (quasi)-particle $\eta\in\lbrace e,m\rbrace$ with spin $\sigma$ and we have assumed the spin branches to be independent, i.e., we work in the limit of no SOC, or very weak SOC relative to the spin-splitting, rendering the spin degree of freedom the correct quantum number within the Boltzmann equation. Here, $\langle f^\sigma_\eta\rangle$ is the equilibrium distribution function in the reference frame moving with the SAW, which will also depend on the local quasiparticle density $N_\eta(\bm r,t)$. Furthermore, $\tau_\eta$ is the (spin-independent) relaxation time and $\bm v_{\eta\bm k}^\sigma\equiv\partial\omega_{\eta\bm k}^\sigma/\partial \bm k$ is the quasiparticle velocity. 
	
	The main force on the electrons results from the piezoelectric field, $\bm E_{\PZT}$. Furthermore, the mobility of the electrons results in an induced electric field, $\bm E_i$, as a function of the electron density through $\epsilon_0\bm\nabla\cdot(\epsilon(z)\bm E_i)=\rho$, with $\epsilon_0$ the vacuum permittivity and $\epsilon(z)$ the relative permittivity of the environment, i.e., in this case a piezoelectric with relative permittivity $\epsilon_{\mathrm{PZ}}$ below and the vacuum above. The charge density is given by $\rho=e[N_e(\bm r,t)-n]$ inside the altermagnet, where $n$ is the equilibrium electron density. Thus, $\bm E_i=-ie\sum_{\bm q} e^{-i\bm q\cdot \bm r}\bm q\, (N-n) / (q\epsilon_0(1+\epsilon))$, where the local electron density is induced by the fluctuations up to first order by the piezoelectric field. Here we have assumed that the charge is distributed homogeneously along the $z$ axis, which is a valid assumption if the penetration depth of the SAW is much larger than the film thickness \cite{leeDielectricFunctionQuasitwodimensional1983}. Here $\epsilon=(\epsilon_{\mathrm{PZ}}+1)/2$ is the average relative permittivity.
	Thus, we have $\bm F_e = e\bm E$, with $\bm E=\bm E_{\PZT}+\bm E^i$.

	The magnons, on the other hand, are electrically neutral, and thus the main force on them is the lattice deformation \cite{gurevichMagnetizationOscillationsWaves1996}.
	The lattice deformation can be separated into slow and fast components, where slow and fast are defined relative to the spatial modulation of the magnon wave packet with wave number $\bm k$. For $k \gg q$, the lattice deformation acts as an effective force, whereas if  $k \ll q$ the lattice deformation acts as a scattering potential \cite{weilerSpinPumpingCoherent2012}, entering the collision integral in the Boltzmann equation as phonon-magnon scattering. It is not possible to definitely separate these two processes, and a semiclassical magnon wave packet with a wavelength comparable to $\bm q$ is in fact ill-defined \cite{marderCondensedMatterPhysics2010}. However, both the SAW and the magnons have a linear dispersion for the applicable wavevector range, but with different velocities, and thus magnon-phonon hybridization is forbidden. We therefore disregard magnon-phonon hybridization and incorporate the lattice deformation only as an effective force on the short-wavelength magnons. The primary force is the strain modulation of the bond length between atomic moments, modulating the effective exchange interaction \cite{streibMagnonphononInteractionsMagnetic2019} and thus generating an effective force $\bm F=-\nabla_{\bm r} H_{\strain}$, where $H_{\strain}$ is the effective magnon strain Hamiltonian, which we derive in Sec.~V in the supplementary material (SM) \footnote{See Supplemental Material under ``Ancillary files'' for details on the derivation of the spin current in the metallic and insulating altermagnet, the detection of the transverse spin current, the surface acoustic wave, the $z$-component of the electric field, thickness effects and a possible experimental setup. Includes Refs.~\cite{andoElectronicPropertiesTwodimensional1982,kittelQuantumTheorySolids1987,ashcroftSolidStatePhysics1976,ruckriegelAngularMomentumConservation2020,troncosoSpinTransportThick2020,rezendeDiffusiveMagnonicSpin2016,yamanouchiPropagationAmplificationRayleigh1972,weisLithiumNiobateSummary1985,kingViscosityTensorApproach1969,fallMeasurementSurfaceAcoustic2025}.}. 
	
	Having obtained the forces on both the electrons and magnons, we solve the Boltzmann equation to find a DC response. Since the forces have a finite frequency set by the SAW frequency $\omega$ , the time average of the force vanishes and there is no first order DC response to $\bm F$. Following Ref.~\cite{kalameitsevValleyAcoustoelectricEffect2019}, we therefore expand up to second order the distribution function $f_\eta^\sigma=f^\sigma_{\eta0} + f^\sigma_{\eta1} + f^\sigma_{\eta2}$, the local equilibrium distribution function $\langle f_\eta^\sigma\rangle=f^\sigma_{\eta0} + (n^\sigma_{\eta1}+n_{\eta2}^\sigma)\frac{\partial f_\eta^\sigma}{\partial n_\eta}$ and the local electron density $N_\eta(\bm r,t)=f^\sigma_{\eta0} + (n^\sigma_{\eta1}+n_{\eta2}^\sigma)\frac{\partial f_\eta^\sigma}{\partial n_\eta} + \frac12(n^\sigma_{\eta1}+n_{\eta2}^\sigma)\frac{\partial^2 f_\eta^\sigma}{\partial n_\eta^2} $. Here $f_{\eta0}^\sigma$ is the equilibrium distribution function, which will be the Fermi-Dirac distribution, $f_{e0}^\sigma=1/(\exp[(\hbar\omega^\sigma_{e\bm k}-\mu_e)/k_BT]+1)$ for electrons, and the Bose-Einstein distribution $f_{m0}^\sigma=1/(\exp[(\hbar\omega^\sigma_{m\bm k}-\mu_m)/k_BT]-1)$ for magnons. We have parameterized the magnon non-equilibrium distribution through a magnon chemical potential, which is possible at timescales slower than the fast exchange-driven magnon-magnon scattering \cite{cornelissenMagnonSpinTransport2016}. We neglect possible changes in temperature, assuming the SAW power to be too small to induce any heating. Experimental setups will of course induce a small amount of heating, but these thermal effects can most likely be isolated through an analysis of their temperature dependency \cite{zhouCrystalThermalTransport2024}.
	
	We will only sketch the derivation here, and relegate the detailed derivation to Secs.~I and II in the SM. For the electrons, we first calculate the first-order response, and find a current $\jwone{e}$ of quasiparticles with spin $\sigma$ at frequency $\omega$. This current $\jwone{e}$ will induce an electron density $n_{e1}^\sigma$, which generates the induced electric $\bm E_i$, describing the Thomas-Fermi screening of the piezoelectric field. From the obtained current $\jwone{e}$ and the induced field $\bm E_i$ we can now solve for the second-order current, which will have two contributions: a rectified direct current $\jtwo{e}$ and a higher harmonic current $\jwtwo{e}$ at frequency $2\omega$.

	For the magnonic response, we follow the same derivation, with $\bm F=-\nabla_{\bm{r}}H_{\strain}$ as the force on the magnons; details are shown in Secs.~III-V in the SM \cite{Note1} There is no magnon analagoue of the induced electric field $\bm E_i$, and thus no direct screening. There will however be a screening induced by the diffusion of magnons.

	\textit{Spin currents. }
	We work at room temperature, where the thermal energy $k_BT$ is much smaller than the electron Fermi energy, and we can thus consider the electrons to be at zero temperature. The magnons on the other hand, are non-conserved bosonic qausiparticles, and we thus consider the magnons at $T=\SI{300}{K}$. We furthermore continue in the limit of $\omega\tau_\eta\ll1,\bm v_{\eta\bm k}^\sigma\cdot\bm q\ll1$ for both electrons and magnons with either spin $\sigma$, where $\bm v_{\eta\bm k}^\sigma$ is the Fermi velocity for electrons and the thermal group velocity for magnons. This limit can be reasonably achieved given the frequency and velocity mismatch between SAWs and electrons and magnons. 
	
	Because of the spin-split nature of both the magnon and electron dispersion in an altermagnet, this current will be spin-polarized, with an anisotropy following the altermagnetic symmetry. In what follows, we will consider the setup as shown in \cref{fig:setup}, where the SAW is coming at an angle $\theta$ with respect to the altermagnetic crystallographic axis. We assume the N\'{e}el vector to lie along the $y$-axis, such that the spin current along the film normal is also polarized along $y$, as was recently achieved in \ch{RuO2} thin films \cite{liaoSeparationInverseAltermagnetic2024}. Therefore, in the heavy metal layer, the inverse spin Hall effect will generate a charge current and voltage along $x$, if leads are attached \cite{sinovaSpinHallEffects2015}. In addition, the out-of-plane deformation $u_{xz}$ and piezoelectric field $E_z$ will drive a charge current along $z$, but this will be a constant contribution as a function of SAW angle $\theta$, and can therefore be isolated, as discussed in Sec.~VIII in the SM \cite{Note1}.

	In addition, we make use of the fact that $\alpha_e<\hbar^2/2m^*$ and $\alpha_mk_T<v_s$, where $k_T=k_BT/v_s$ is the thermal magnon wave vector, and thus we expand the result up to first order in $\alpha_e$ and $\alpha_m$. This allows us to obtain closed analytical expressions for the currents. In metallic altermagnetic films we find a transverse spin current carried by electrons,
	\begin{equation}
		j_{e;s(0)}^{z} = - \frac{\tau_e}{\hbar^2}  \frac{ q\sigma_e^0} {\omega } \frac{|E|^2}{1+\zeta_e^2} \alpha_e\cos\theta.
		\label{eq:jsH}
	\end{equation}
	Here $\sigma_{e}^0$ is the spin-independent electrical conductivity and $\zeta_e \equiv v^* q/\omega$ is a dimensionless parameter describing the induced field response, where $v^{*}=\sigma_{e}^0 d/\epsilon_0(1+\epsilon)$ is the velocity of charge spreading. We note that $\sigma_e^0$ is also anisotropic due to the anisotropic band structure, but this gives rises to $O(\alpha_e^2)$ corrections to \cref{eq:jsH}. Alternatively, the anisotropic conductance could be extracted experimentally through electrical measurements.
	
	In insulating altermagnetic films we find the transverse spin current carried by magnons as
	\begin{equation}
		j_{m;s(0)}^{z} = - \frac{\tau_m}{\hbar^2} \frac{\partial \mu_m}{\partial n_m} \frac{ q\sigma_{m}^0} {\omega } \frac{|F_{\mathrm{eff}}|^2}{1+\chi_m} \alpha_m h(\theta)\cos\theta \label{eq:jsH-magnon},
	\end{equation}
	where $F_{\mathrm{eff}}\equiv (k_B T)^{2}\beta_m u_0q^2/(\pi v_s)^{3/2}$ is the effective force, with $\beta_m$ the magnetic Grüneisen parameter \cite{bloch10LawVolume1966} and $u_0$ the deformation amplitude. $h(\theta)\equiv\htheta$ incorporates an additional eight-fold rotational symmetry due to the fact that the coupling between a strain component $u_{\alpha\beta}$ and an exchange bond along $\bm r$ goes as $(\hat\alpha\cdot\hat r)(\hat\beta\cdot\hat r)$. Therefore we capture the four-fold rotational symmetry of the crystal lattice, becoming an eight-fold rotational symmetry in the second-order response. Here, $\varrho$ is a dimensionless piezoelectric coupling factor, which relates the strain components $u_{xy}$ and $u_{yz}$. There is no magnon analogue of the induced field, but diffusion is allowed and thus $\chi_m=q^2\sigma_m^0 (\partial \mu_m/\partial n_m) / \omega$. Finally, $\partial \mu_m/\partial n_m=\pi^2 v_s^3/(k_BT)^2$ is the change in magnon chemical potential $\mu_m$ as the magnon density $n_m$ is varied.

	In addition to the second order DC response, the setup as described here also generates higher-order harmonics, which could be measured through a lock-in technique. The first higher harmonic is at frequency $2\omega$, and generates a transverse electron spin current, $j_{e;s(2\omega)}^{z}=\frac{1-\zeta_e^2}{1+\zeta_e^2}j_{e;s(0)}^{z}$ and magnon spin current $j_{m;s(2\omega)}^{z}=j_{m;s(0)}^{z}$.
	
	\begin{figure}
		\centering
		\includegraphics{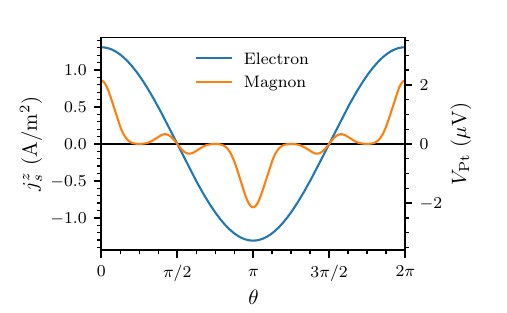}
		\caption{The transverse spin current (left axis) and the resulting voltage in the Platinum (right axis) as a function of the angle of the SAW to the crystallographic axis. Results shown for altermagnetic films with thickness $d=\thickness$ and SAW frequency $\omega/2\pi=\SAWomega$. \label{fig:theta}}
	\end{figure}

	\textit{Results. }
	Having obtained closed expressions for the transverse spin current, we now calculate the expected spin current, assuming the metallic thin film to be \ch{RuO2}. The electron band structure of the magnetic state has been calculated by \emph{ab initio} calculations \cite{ahnAntiferromagnetism$mathrmRuO_2$$d$wave2019,gonzalez-hernandezEfficientElectricalSpin2021,smejkalConventionalFerromagnetismAntiferromagnetism2022,smejkalEmergingResearchLandscape2022}, from which we find $m^*=\meff m_e$, $\alpha_e=\alphae \hbar^2/2m^*$, where $m_e$ is the electron mass. In thin films, the electron carrier density and conductivity have been measured to be $n=\density$ and $\sigma_e^0=\sigmae$ \cite{tschirnerSaturationAnomalousHall2023}, such that $\tau_e=\sigma_e^0m^*/ne^2\approx\taue$ in the Drude approximation. 
	
	There are currently no known insulating altermagnets with an observable magnon splitting \cite{moranoAbsenceAltermagneticMagnon2025}, and we therefore assume a rutile-based insulating altermagnet, taking the exchange parameters calculated for \ch{RuO2} \cite{smejkalChiralMagnonsAltermagnetic2023}, such that $v_s=\vsmagnon$  and $\alpha_m=\alpham$; see also Sec.~IV in the SM \cite{Note1}. To estimate the magnon lifetime, we assume a damping rate of $\gamma^{-1}=\alphag$ and find for the thermal magnons $\tau_m=\gamma \hbar/(2k_BT)\approx\taum$. Additionally, we assume that $\sigma_m\times e^2=\sigmam$, in accordance with measurements of the magnon conductivity in insulating antiferromagnets \cite{lebrunTunableLongdistanceSpin2018}.
	We take the typical value of $\beta_m\approx-10/3$  for the magnetic Gr\"uneisen parameter, which is valid in a wide range of materials \cite{bloch10LawVolume1966,samaraEffectPressureNeel1969,streibMagnonphononInteractionsMagnetic2019}. 
	
	We assume the piezoelectric substrate to be \ch{LiNbO3}, and give further details of the BG SAW in Sec.~VII in the SM \cite{Note1}. Important here is the sound velocity, $s\approx\SAWspeed$, the proportionality constant $c_{PZ}\approx\utophizero$, the piezoelectric coupling factor $\varrho\approx\piezovarrho$ and the SAW intensity given by $I_0=I_\omega \omega$, where $I_\omega\approx\Izeropropto$. 
	At $\omega/2\pi=\SAWomega$, this corresponds to an electric field of $\Ezero$ and an average deformation of the unit cell of $\deformation a$.

	In what follows, we show the transverse spin current and the resulting voltage in the heavy metal layer. We assume the heavy metal layer to be \thicknessplatinum\ thick Platinum, with Hall angle $\theta_{H}=\hallangle$ \cite{wangScalingSpinHall2014}, forming a Hall bar with sides of $l_x=l_y=\leadlength$. The induced open circuit voltage along $\hat{\bm{x}}$ by a transverse spin current $j_s$ is then $V_{\mathrm{Pt}}= \theta_H j_s A R$, where $R=\rho l_x / (l_y d)$, with $\rho=\rhoplatinum$ the resistivity of Platinum \cite{wangScalingSpinHall2014} and $A=l_xl_y$; see Sec.~VI in the SM \cite{Note1}.
	
	We show the angle dependency of the transverse spin current in \cref{fig:theta}, as the SAW is rotated around the altermagnetic sample, showing the spin current generated in a metallic and insulating thin film with thickness $d=\thickness$.
	We first observe that the transverse spin current displays the same $d$-wave symmetry as the altermagnetic crystal, such that the transverse spin current is zero for SAWs travelling along the high-symmetry planes $\theta=\pm\pi/2$ and is maximized for $\theta\in [0,\pi]$. The magnon spin current has an additional eight-fold symmetry, related to the four-fold rotational symmetry of the nearest-neighbor bonds in rutile compounds, which becomes an eight-fold symmetry in second response.

\begin{figure}
		\centering
		\includegraphics{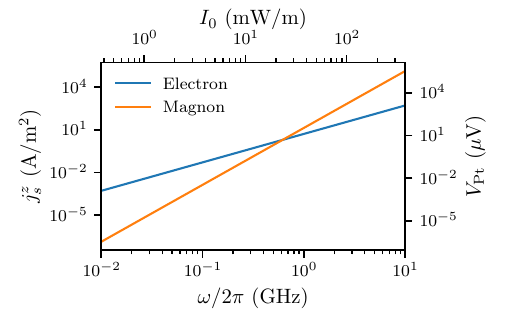}
		\caption{The transverse spin current, as a function of SAW frequency $\omega$. The top axis shows the SAW intensity $I_0$. \label{fig:omega}}
	\end{figure}

	We also observe from \cref{fig:omega} that the magnonic and electronic contributions are of comparable amplitude, which \emph{a priori} might seem surprising. This can however be attributed to the fact that the electron system is highly efficient in screening the induced first-order current, parametrized by $\zeta_e$. For the parameters used here, $\zeta_e\approx\zetaelectron$, which thus suppresses the second-order electron current by eight orders of magnitude. For the magnons on the other hand, no  screening through charge spreading exists and only the diffusion screens the first-order current. This is parameterized by $\chi_m$, but $\chi_m\approx\chimagnon$ for the parameters used here. The electron screening is even stronger for thicker metallic films, as discussed in Sec.~IX in the SM \cite{Note1}.
	
	In \cref{fig:omega} we show the transverse spin current for electrons and magnons as a function of the SAW frequency, $\omega$. The top axis shows the SAW intensity, which is linear in $\omega$. We observe a different scaling as $\omega$ for the induced magnon ($\propto\omega^4$) and electron ($\propto\omega^2$) current. 
	This difference can be traced to the effective force on the magnons, which follows from the strain modulating the exchange bond strengths. This force is proportional to $q^2$, as opposed to the electric field, which goes as $q$. (Note that $q=\omega s$.) 

	\textit{Conclusion and discussion. }
	We have shown that it is possible to excite electron and magnon spin currents through a surface acoustic wave in metallic and insulating altermagnetic thin films, offering broad material applicability and a novel way to detect the spin splitter effect in insulating altermagnetic thin film. For realistic parameters, we have predicted that the spin currents can be measured through the inverse spin Hall effect in a heavy metal layer. Finally, we have shown that the SAW frequency can be used to tune electron and magnon responses. 
	
	Other orientations of the altermagnetic crystal structure are also possible, which would generate spin currents along different crystallographic directions and with different spin polarization directions. The setup proposed here, which corresponds to \ch{RuO2} grown along the (100) orientation \cite{liaoSeparationInverseAltermagnetic2024}, was chosen because it generates an inverse spin Hall voltage in the heavy metal along the $x$ axis, but spin currents can still be generated in other crystal orientations.
	
	In addition to the acoustic spin splitter discussed here, there will be additional contributions to the voltage along $x$ in the heavy metal layer: (i) direct generation of a charge current along $x$ in the heavy metal layer through the conventional acousto-electric effect \cite{parmenterAcoustoElectricEffect1953, weinreichObservationAcoustoelectricEffect1957}; (ii) generation of a transverse spin current in the metallic altermagnet through the relativistic spin Hall effect (SHE); (iii) a temperature gradient in the altermagnetic thin film, which will drive an additional transverse spin current through the spin splitter Nernst effect in $d$-wave altermagnets \cite{cuiEfficientSpinSeebeck2023,yiSpinSplittingNernst2025}. To fully isolate the acoustic spin splitter effect we propose an experimental setup in Sec.~X in the SM \cite{Note1}. We show that by rotating the angle of the SAW $\theta$, studying the frequency dependency, and controlling the direction of the N\'eel vector, one can isolate the acoustic spin splitter effect. We believe the most feasible platform to conclusively isolate the acoustic spin splitter will be an insulating altermagnet, preferably with control over the the direction of the N\'eel vector.
	
	
	Under experimental conditions, an altermagnetic thin film is most likely composed of multiple domains \cite{aminNanoscaleImagingControl2024}. This could potentially reduce the signal, but recent experiments indicate that high-enough quality samples retain a finite anomalous Hall response, which can be attributed to the altermagnetic splitting \cite{jeongMetallicityAnomalousHall2025, reichlovaObservationSpontaneousAnomalous2024}.


	We envision that SAWs offer an enticing path platform to not only generate spin currents in altermagnets, but also to control the  altermagnetic symmetries, moving beyond the static strain that has thus far been used \cite{chakrabortyStraininducedPhaseTransition2024,karettaStraincontrolledWaveTransition2025, belashchenkoGiantStrainInducedSpin2025, liExtremeStrainControlled2025, zhouManipulationAltermagneticOrder2025}. For example, the dynamical $u_{xy}$ strain generated by a BG SAW will induce a dynamical spin splitting in \ch{MnTe}, a $g$-wave altermagnetic semiconductor \cite{belashchenkoGiantStrainInducedSpin2025}, or in \ch{CrSb}, a metallic $g$-wave altermagnet \cite{karettaStraincontrolledWaveTransition2025}. Importantly, this is a dynamical effect, i.e., $\propto e^{-i\omega t}$, and the induced spin splitting will switch sign over a full time period of the SAW.

	The data that support the findings of this article are openly available \cite{gunninkSurfaceAcousticWave2026}.

	\begin{acknowledgments}
		We thank Mathias Weiler for stimulating discussions. This work is in part funded by the Deutsche Forschungsgemeinschaft (DFG, German Research Foundation) -- Project No.~504261060 (Emmy Noether Programme), and TRR~173-268565370 (project A03 and B13). P.~G. acknowledges financial support from an Alexander von Humboldt postdoctoral fellowship.
	\end{acknowledgments}


\begin{thebibliography}{61}%
		\makeatletter
		\providecommand \@ifxundefined [1]{%
			\@ifx{#1\undefined}
		}%
		\providecommand \@ifnum [1]{%
			\ifnum #1\expandafter \@firstoftwo
			\else \expandafter \@secondoftwo
			\fi
		}%
		\providecommand \@ifx [1]{%
			\ifx #1\expandafter \@firstoftwo
			\else \expandafter \@secondoftwo
			\fi
		}%
		\providecommand \natexlab [1]{#1}%
		\providecommand \enquote  [1]{``#1''}%
		\providecommand \bibnamefont  [1]{#1}%
		\providecommand \bibfnamefont [1]{#1}%
		\providecommand \citenamefont [1]{#1}%
		\providecommand \href@noop [0]{\@secondoftwo}%
		\providecommand \href [0]{\begingroup \@sanitize@url \@href}%
		\providecommand \@href[1]{\@@startlink{#1}\@@href}%
		\providecommand \@@href[1]{\endgroup#1\@@endlink}%
		\providecommand \@sanitize@url [0]{\catcode `\\12\catcode `\$12\catcode
			`\&12\catcode `\#12\catcode `\^12\catcode `\_12\catcode `\%12\relax}%
		\providecommand \@@startlink[1]{}%
		\providecommand \@@endlink[0]{}%
		\providecommand \url  [0]{\begingroup\@sanitize@url \@url }%
		\providecommand \@url [1]{\endgroup\@href {#1}{\urlprefix }}%
		\providecommand \urlprefix  [0]{URL }%
		\providecommand \Eprint [0]{\href }%
		\providecommand \doibase [0]{https://doi.org/}%
		\providecommand \selectlanguage [0]{\@gobble}%
		\providecommand \bibinfo  [0]{\@secondoftwo}%
		\providecommand \bibfield  [0]{\@secondoftwo}%
		\providecommand \translation [1]{[#1]}%
		\providecommand \BibitemOpen [0]{}%
		\providecommand \bibitemStop [0]{}%
		\providecommand \bibitemNoStop [0]{.\EOS\space}%
		\providecommand \EOS [0]{\spacefactor3000\relax}%
		\providecommand \BibitemShut  [1]{\csname bibitem#1\endcsname}%
		\let\auto@bib@innerbib\@empty
		\bibitem [{\citenamefont {Sinova}\ \emph {et~al.}(2015)\citenamefont {Sinova},
			\citenamefont {Valenzuela}, \citenamefont {Wunderlich}, \citenamefont
			{Back},\ and\ \citenamefont {Jungwirth}}]{sinovaSpinHallEffects2015}%
		\BibitemOpen
		\bibfield  {author} {\bibinfo {author} {\bibfnamefont {J.}~\bibnamefont
				{Sinova}}, \bibinfo {author} {\bibfnamefont {S.~O.}\ \bibnamefont
				{Valenzuela}}, \bibinfo {author} {\bibfnamefont {J.}~\bibnamefont
				{Wunderlich}}, \bibinfo {author} {\bibfnamefont {C.~H.}\ \bibnamefont
				{Back}},\ and\ \bibinfo {author} {\bibfnamefont {T.}~\bibnamefont
				{Jungwirth}},\ }\bibfield  {title} {\bibinfo {title} {Spin {{Hall}}
				effects},\ }\href {https://doi.org/10.1103/RevModPhys.87.1213} {\bibfield
			{journal} {\bibinfo  {journal} {Reviews of Modern Physics}\ }\textbf
			{\bibinfo {volume} {87}},\ \bibinfo {pages} {1213} (\bibinfo {year}
			{2015})}\BibitemShut {NoStop}%
		\bibitem [{\citenamefont {Manchon}\ \emph {et~al.}(2019)\citenamefont
			{Manchon}, \citenamefont {{\v Z}elezn{\'y}}, \citenamefont {Miron},
			\citenamefont {Jungwirth}, \citenamefont {Sinova}, \citenamefont {Thiaville},
			\citenamefont {Garello},\ and\ \citenamefont
			{Gambardella}}]{manchonCurrentinducedSpinorbitTorques2019}%
		\BibitemOpen
		\bibfield  {author} {\bibinfo {author} {\bibfnamefont {A.}~\bibnamefont
				{Manchon}}, \bibinfo {author} {\bibfnamefont {J.}~\bibnamefont {{\v
						Z}elezn{\'y}}}, \bibinfo {author} {\bibfnamefont {I.~M.}\ \bibnamefont
				{Miron}}, \bibinfo {author} {\bibfnamefont {T.}~\bibnamefont {Jungwirth}},
			\bibinfo {author} {\bibfnamefont {J.}~\bibnamefont {Sinova}}, \bibinfo
			{author} {\bibfnamefont {A.}~\bibnamefont {Thiaville}}, \bibinfo {author}
			{\bibfnamefont {K.}~\bibnamefont {Garello}},\ and\ \bibinfo {author}
			{\bibfnamefont {P.}~\bibnamefont {Gambardella}},\ }\bibfield  {title}
		{\bibinfo {title} {Current-induced spin-orbit torques in ferromagnetic and
				antiferromagnetic systems},\ }\href
		{https://doi.org/10.1103/RevModPhys.91.035004} {\bibfield  {journal}
			{\bibinfo  {journal} {Reviews of Modern Physics}\ }\textbf {\bibinfo {volume}
				{91}},\ \bibinfo {pages} {035004} (\bibinfo {year} {2019})}\BibitemShut
		{NoStop}%
		\bibitem [{\citenamefont {{\v S}mejkal}\ \emph
			{et~al.}(2022{\natexlab{a}})\citenamefont {{\v S}mejkal}, \citenamefont
			{Sinova},\ and\ \citenamefont
			{Jungwirth}}]{smejkalConventionalFerromagnetismAntiferromagnetism2022}%
		\BibitemOpen
		\bibfield  {author} {\bibinfo {author} {\bibfnamefont {L.}~\bibnamefont {{\v
						S}mejkal}}, \bibinfo {author} {\bibfnamefont {J.}~\bibnamefont {Sinova}},\
			and\ \bibinfo {author} {\bibfnamefont {T.}~\bibnamefont {Jungwirth}},\
		}\bibfield  {title} {\bibinfo {title} {Beyond {{Conventional Ferromagnetism}}
				and {{Antiferromagnetism}}: {{A Phase}} with {{Nonrelativistic Spin}} and
				{{Crystal Rotation Symmetry}}},\ }\href
		{https://doi.org/10.1103/PhysRevX.12.031042} {\bibfield  {journal} {\bibinfo
				{journal} {Physical Review X}\ }\textbf {\bibinfo {volume} {12}},\ \bibinfo
			{pages} {031042} (\bibinfo {year} {2022}{\natexlab{a}})}\BibitemShut
		{NoStop}%
		\bibitem [{\citenamefont {{\v S}mejkal}\ \emph {et~al.}(2020)\citenamefont {{\v
					S}mejkal}, \citenamefont {{Gonz{\'a}lez-Hern{\'a}ndez}}, \citenamefont
			{Jungwirth},\ and\ \citenamefont
			{Sinova}}]{smejkalCrystalTimereversalSymmetry2020}%
		\BibitemOpen
		\bibfield  {author} {\bibinfo {author} {\bibfnamefont {L.}~\bibnamefont {{\v
						S}mejkal}}, \bibinfo {author} {\bibfnamefont {R.}~\bibnamefont
				{{Gonz{\'a}lez-Hern{\'a}ndez}}}, \bibinfo {author} {\bibfnamefont
				{T.}~\bibnamefont {Jungwirth}},\ and\ \bibinfo {author} {\bibfnamefont
				{J.}~\bibnamefont {Sinova}},\ }\bibfield  {title} {\bibinfo {title} {Crystal
				time-reversal symmetry breaking and spontaneous {{Hall}} effect in collinear
				antiferromagnets},\ }\href {https://doi.org/10.1126/sciadv.aaz8809}
		{\bibfield  {journal} {\bibinfo  {journal} {Science Advances}\ }\textbf
			{\bibinfo {volume} {6}},\ \bibinfo {pages} {eaaz8809} (\bibinfo {year}
			{2020})}\BibitemShut {NoStop}%
		\bibitem [{\citenamefont {{\v S}mejkal}\ \emph
			{et~al.}(2022{\natexlab{b}})\citenamefont {{\v S}mejkal}, \citenamefont
			{Sinova},\ and\ \citenamefont
			{Jungwirth}}]{smejkalEmergingResearchLandscape2022}%
		\BibitemOpen
		\bibfield  {author} {\bibinfo {author} {\bibfnamefont {L.}~\bibnamefont {{\v
						S}mejkal}}, \bibinfo {author} {\bibfnamefont {J.}~\bibnamefont {Sinova}},\
			and\ \bibinfo {author} {\bibfnamefont {T.}~\bibnamefont {Jungwirth}},\
		}\bibfield  {title} {\bibinfo {title} {Emerging {{Research Landscape}} of
				{{Altermagnetism}}},\ }\href {https://doi.org/10.1103/PhysRevX.12.040501}
		{\bibfield  {journal} {\bibinfo  {journal} {Physical Review X}\ }\textbf
			{\bibinfo {volume} {12}},\ \bibinfo {pages} {040501} (\bibinfo {year}
			{2022}{\natexlab{b}})}\BibitemShut {NoStop}%
		\bibitem [{\citenamefont {Naka}\ \emph {et~al.}(2019)\citenamefont {Naka},
			\citenamefont {Hayami}, \citenamefont {Kusunose}, \citenamefont {Yanagi},
			\citenamefont {Motome},\ and\ \citenamefont
			{Seo}}]{nakaSpinCurrentGeneration2019}%
		\BibitemOpen
		\bibfield  {author} {\bibinfo {author} {\bibfnamefont {M.}~\bibnamefont
				{Naka}}, \bibinfo {author} {\bibfnamefont {S.}~\bibnamefont {Hayami}},
			\bibinfo {author} {\bibfnamefont {H.}~\bibnamefont {Kusunose}}, \bibinfo
			{author} {\bibfnamefont {Y.}~\bibnamefont {Yanagi}}, \bibinfo {author}
			{\bibfnamefont {Y.}~\bibnamefont {Motome}},\ and\ \bibinfo {author}
			{\bibfnamefont {H.}~\bibnamefont {Seo}},\ }\bibfield  {title} {\bibinfo
			{title} {Spin current generation in organic antiferromagnets},\ }\href
		{https://doi.org/10.1038/s41467-019-12229-y} {\bibfield  {journal} {\bibinfo
				{journal} {Nature Communications}\ }\textbf {\bibinfo {volume} {10}},\
			\bibinfo {pages} {4305} (\bibinfo {year} {2019})}\BibitemShut {NoStop}%
		\bibitem [{\citenamefont {{Gonz{\'a}lez-Hern{\'a}ndez}}\ \emph
			{et~al.}(2021)\citenamefont {{Gonz{\'a}lez-Hern{\'a}ndez}}, \citenamefont
			{{\v S}mejkal}, \citenamefont {V{\'y}born{\'y}}, \citenamefont {Yahagi},
			\citenamefont {Sinova}, \citenamefont {Jungwirth},\ and\ \citenamefont {{\v
					Z}elezn{\'y}}}]{gonzalez-hernandezEfficientElectricalSpin2021}%
		\BibitemOpen
		\bibfield  {author} {\bibinfo {author} {\bibfnamefont {R.}~\bibnamefont
				{{Gonz{\'a}lez-Hern{\'a}ndez}}}, \bibinfo {author} {\bibfnamefont
				{L.}~\bibnamefont {{\v S}mejkal}}, \bibinfo {author} {\bibfnamefont
				{K.}~\bibnamefont {V{\'y}born{\'y}}}, \bibinfo {author} {\bibfnamefont
				{Y.}~\bibnamefont {Yahagi}}, \bibinfo {author} {\bibfnamefont
				{J.}~\bibnamefont {Sinova}}, \bibinfo {author} {\bibfnamefont
				{T.}~\bibnamefont {Jungwirth}},\ and\ \bibinfo {author} {\bibfnamefont
				{J.}~\bibnamefont {{\v Z}elezn{\'y}}},\ }\bibfield  {title} {\bibinfo {title}
			{Efficient {{Electrical Spin Splitter Based}} on {{Nonrelativistic Collinear
						Antiferromagnetism}}},\ }\href
		{https://doi.org/10.1103/PhysRevLett.126.127701} {\bibfield  {journal}
			{\bibinfo  {journal} {Physical Review Letters}\ }\textbf {\bibinfo {volume}
				{126}},\ \bibinfo {pages} {127701} (\bibinfo {year} {2021})}\BibitemShut
		{NoStop}%
		\bibitem [{\citenamefont {Sourounis}\ and\ \citenamefont
			{Manchon}(2024)}]{sourounisEfficientGenerationSpin2024}%
		\BibitemOpen
		\bibfield  {author} {\bibinfo {author} {\bibfnamefont {K.}~\bibnamefont
				{Sourounis}}\ and\ \bibinfo {author} {\bibfnamefont {A.}~\bibnamefont
				{Manchon}},\ }\href {https://doi.org/10.48550/arXiv.2411.14803} {\bibinfo
			{title} {Efficient {{Generation}} of {{Spin Currents}} in {{Altermagnets}}
				via {{Magnon Drag}}}} (\bibinfo {year} {2024}),\ \Eprint
		{https://arxiv.org/abs/2411.14803} {arXiv:2411.14803 [cond-mat]} \BibitemShut
		{NoStop}%
		\bibitem [{\citenamefont
			{Parmenter}(1953)}]{parmenterAcoustoElectricEffect1953}%
		\BibitemOpen
		\bibfield  {author} {\bibinfo {author} {\bibfnamefont {R.~H.}\ \bibnamefont
				{Parmenter}},\ }\bibfield  {title} {\bibinfo {title} {The {{Acousto-Electric
						Effect}}},\ }\href {https://doi.org/10.1103/PhysRev.89.990} {\bibfield
			{journal} {\bibinfo  {journal} {Physical Review}\ }\textbf {\bibinfo {volume}
				{89}},\ \bibinfo {pages} {990} (\bibinfo {year} {1953})}\BibitemShut
		{NoStop}%
		\bibitem [{\citenamefont {Weinreich}\ and\ \citenamefont
			{White}(1957)}]{weinreichObservationAcoustoelectricEffect1957}%
		\BibitemOpen
		\bibfield  {author} {\bibinfo {author} {\bibfnamefont {G.}~\bibnamefont
				{Weinreich}}\ and\ \bibinfo {author} {\bibfnamefont {H.~G.}\ \bibnamefont
				{White}},\ }\bibfield  {title} {\bibinfo {title} {Observation of the
				{{Acoustoelectric Effect}}},\ }\href
		{https://doi.org/10.1103/PhysRev.106.1104} {\bibfield  {journal} {\bibinfo
				{journal} {Physical Review}\ }\textbf {\bibinfo {volume} {106}},\ \bibinfo
			{pages} {1104} (\bibinfo {year} {1957})}\BibitemShut {NoStop}%
		\bibitem [{\citenamefont {Kalameitsev}\ \emph {et~al.}(2019)\citenamefont
			{Kalameitsev}, \citenamefont {Kovalev},\ and\ \citenamefont
			{Savenko}}]{kalameitsevValleyAcoustoelectricEffect2019}%
		\BibitemOpen
		\bibfield  {author} {\bibinfo {author} {\bibfnamefont {A.~V.}\ \bibnamefont
				{Kalameitsev}}, \bibinfo {author} {\bibfnamefont {V.~M.}\ \bibnamefont
				{Kovalev}},\ and\ \bibinfo {author} {\bibfnamefont {I.~G.}\ \bibnamefont
				{Savenko}},\ }\bibfield  {title} {\bibinfo {title} {Valley {{Acoustoelectric
						Effect}}},\ }\href {https://doi.org/10.1103/PhysRevLett.122.256801}
		{\bibfield  {journal} {\bibinfo  {journal} {Physical Review Letters}\
			}\textbf {\bibinfo {volume} {122}},\ \bibinfo {pages} {256801} (\bibinfo
			{year} {2019})}\BibitemShut {NoStop}%
		\bibitem [{\citenamefont {Sonowal}\ \emph {et~al.}(2020)\citenamefont
			{Sonowal}, \citenamefont {Kalameitsev}, \citenamefont {Kovalev},\ and\
			\citenamefont {Savenko}}]{sonowalAcoustoelectricEffectTwodimensional2020}%
		\BibitemOpen
		\bibfield  {author} {\bibinfo {author} {\bibfnamefont {K.}~\bibnamefont
				{Sonowal}}, \bibinfo {author} {\bibfnamefont {A.~V.}\ \bibnamefont
				{Kalameitsev}}, \bibinfo {author} {\bibfnamefont {V.~M.}\ \bibnamefont
				{Kovalev}},\ and\ \bibinfo {author} {\bibfnamefont {I.~G.}\ \bibnamefont
				{Savenko}},\ }\bibfield  {title} {\bibinfo {title} {Acoustoelectric effect in
				two-dimensional {{Dirac}} materials exposed to {{Rayleigh}} surface acoustic
				waves},\ }\href {https://doi.org/10.1103/PhysRevB.102.235405} {\bibfield
			{journal} {\bibinfo  {journal} {Physical Review B}\ }\textbf {\bibinfo
				{volume} {102}},\ \bibinfo {pages} {235405} (\bibinfo {year}
			{2020})}\BibitemShut {NoStop}%
		\bibitem [{\citenamefont {Bhalla}\ \emph {et~al.}(2022)\citenamefont {Bhalla},
			\citenamefont {Vignale},\ and\ \citenamefont
			{Rostami}}]{bhallaPseudogaugeFieldDriven2022}%
		\BibitemOpen
		\bibfield  {author} {\bibinfo {author} {\bibfnamefont {P.}~\bibnamefont
				{Bhalla}}, \bibinfo {author} {\bibfnamefont {G.}~\bibnamefont {Vignale}},\
			and\ \bibinfo {author} {\bibfnamefont {H.}~\bibnamefont {Rostami}},\
		}\bibfield  {title} {\bibinfo {title} {Pseudogauge field driven
				acoustoelectric current in two-dimensional hexagonal {{Dirac}} materials},\
		}\href {https://doi.org/10.1103/PhysRevB.105.125407} {\bibfield  {journal}
			{\bibinfo  {journal} {Physical Review B}\ }\textbf {\bibinfo {volume}
				{105}},\ \bibinfo {pages} {125407} (\bibinfo {year} {2022})}\BibitemShut
		{NoStop}%
		\bibitem [{\citenamefont {Talyanskii}\ \emph {et~al.}(1997)\citenamefont
			{Talyanskii}, \citenamefont {Shilton}, \citenamefont {Pepper}, \citenamefont
			{Smith}, \citenamefont {Ford}, \citenamefont {Linfield}, \citenamefont
			{Ritchie},\ and\ \citenamefont
			{Jones}}]{talyanskiiSingleelectronTransportOnedimensional1997}%
		\BibitemOpen
		\bibfield  {author} {\bibinfo {author} {\bibfnamefont {V.~I.}\ \bibnamefont
				{Talyanskii}}, \bibinfo {author} {\bibfnamefont {J.~M.}\ \bibnamefont
				{Shilton}}, \bibinfo {author} {\bibfnamefont {M.}~\bibnamefont {Pepper}},
			\bibinfo {author} {\bibfnamefont {C.~G.}\ \bibnamefont {Smith}}, \bibinfo
			{author} {\bibfnamefont {C.~J.~B.}\ \bibnamefont {Ford}}, \bibinfo {author}
			{\bibfnamefont {E.~H.}\ \bibnamefont {Linfield}}, \bibinfo {author}
			{\bibfnamefont {D.~A.}\ \bibnamefont {Ritchie}},\ and\ \bibinfo {author}
			{\bibfnamefont {G.~A.~C.}\ \bibnamefont {Jones}},\ }\bibfield  {title}
		{\bibinfo {title} {Single-electron transport in a one-dimensional channel by
				high-frequency surface acoustic waves},\ }\href
		{https://doi.org/10.1103/PhysRevB.56.15180} {\bibfield  {journal} {\bibinfo
				{journal} {Physical Review B}\ }\textbf {\bibinfo {volume} {56}},\ \bibinfo
			{pages} {15180} (\bibinfo {year} {1997})}\BibitemShut {NoStop}%
		\bibitem [{\citenamefont {Uchida}\ \emph
			{et~al.}(2011{\natexlab{a}})\citenamefont {Uchida}, \citenamefont {Adachi},
			\citenamefont {An}, \citenamefont {Ota}, \citenamefont {Toda}, \citenamefont
			{Hillebrands}, \citenamefont {Maekawa},\ and\ \citenamefont
			{Saitoh}}]{uchidaLongrangeSpinSeebeck2011}%
		\BibitemOpen
		\bibfield  {author} {\bibinfo {author} {\bibfnamefont {K.}~\bibnamefont
				{Uchida}}, \bibinfo {author} {\bibfnamefont {H.}~\bibnamefont {Adachi}},
			\bibinfo {author} {\bibfnamefont {T.}~\bibnamefont {An}}, \bibinfo {author}
			{\bibfnamefont {T.}~\bibnamefont {Ota}}, \bibinfo {author} {\bibfnamefont
				{M.}~\bibnamefont {Toda}}, \bibinfo {author} {\bibfnamefont {B.}~\bibnamefont
				{Hillebrands}}, \bibinfo {author} {\bibfnamefont {S.}~\bibnamefont
				{Maekawa}},\ and\ \bibinfo {author} {\bibfnamefont {E.}~\bibnamefont
				{Saitoh}},\ }\bibfield  {title} {\bibinfo {title} {Long-range spin
				{{Seebeck}} effect and acoustic spin pumping},\ }\href
		{https://doi.org/10.1038/nmat3099} {\bibfield  {journal} {\bibinfo  {journal}
				{Nature Materials}\ }\textbf {\bibinfo {volume} {10}},\ \bibinfo {pages}
			{737} (\bibinfo {year} {2011}{\natexlab{a}})}\BibitemShut {NoStop}%
		\bibitem [{\citenamefont {Uchida}\ \emph {et~al.}(2012)\citenamefont {Uchida},
			\citenamefont {Adachi}, \citenamefont {An}, \citenamefont {Nakayama},
			\citenamefont {Toda}, \citenamefont {Hillebrands}, \citenamefont {Maekawa},\
			and\ \citenamefont {Saitoh}}]{uchidaAcousticSpinPumping2012}%
		\BibitemOpen
		\bibfield  {author} {\bibinfo {author} {\bibfnamefont {K.}~\bibnamefont
				{Uchida}}, \bibinfo {author} {\bibfnamefont {H.}~\bibnamefont {Adachi}},
			\bibinfo {author} {\bibfnamefont {T.}~\bibnamefont {An}}, \bibinfo {author}
			{\bibfnamefont {H.}~\bibnamefont {Nakayama}}, \bibinfo {author}
			{\bibfnamefont {M.}~\bibnamefont {Toda}}, \bibinfo {author} {\bibfnamefont
				{B.}~\bibnamefont {Hillebrands}}, \bibinfo {author} {\bibfnamefont
				{S.}~\bibnamefont {Maekawa}},\ and\ \bibinfo {author} {\bibfnamefont
				{E.}~\bibnamefont {Saitoh}},\ }\bibfield  {title} {\bibinfo {title} {Acoustic
				spin pumping: {{Direct}} generation of spin currents from sound waves in
				{{$\mathrm{Pt}$/$\mathrm{Y_3Fe_5O_{12}}$}}hybrid structures},\ }\href
		{https://doi.org/10.1063/1.3688332} {\bibfield  {journal} {\bibinfo
				{journal} {Journal of Applied Physics}\ }\textbf {\bibinfo {volume} {111}},\
			\bibinfo {pages} {053903} (\bibinfo {year} {2012})}\BibitemShut {NoStop}%
		\bibitem [{\citenamefont {Adachi}\ \emph {et~al.}(2013)\citenamefont {Adachi},
			\citenamefont {Uchida}, \citenamefont {Saitoh},\ and\ \citenamefont
			{Maekawa}}]{adachiTheorySpinSeebeck2013a}%
		\BibitemOpen
		\bibfield  {author} {\bibinfo {author} {\bibfnamefont {H.}~\bibnamefont
				{Adachi}}, \bibinfo {author} {\bibfnamefont {K.-i.}\ \bibnamefont {Uchida}},
			\bibinfo {author} {\bibfnamefont {E.}~\bibnamefont {Saitoh}},\ and\ \bibinfo
			{author} {\bibfnamefont {S.}~\bibnamefont {Maekawa}},\ }\bibfield  {title}
		{\bibinfo {title} {Theory of the spin {{Seebeck}} effect},\ }\href
		{https://doi.org/10.1088/0034-4885/76/3/036501} {\bibfield  {journal}
			{\bibinfo  {journal} {Reports on Progress in Physics}\ }\textbf {\bibinfo
				{volume} {76}},\ \bibinfo {pages} {036501} (\bibinfo {year}
			{2013})}\BibitemShut {NoStop}%
		\bibitem [{\citenamefont {Uchida}\ \emph
			{et~al.}(2011{\natexlab{b}})\citenamefont {Uchida}, \citenamefont {An},
			\citenamefont {Kajiwara}, \citenamefont {Toda},\ and\ \citenamefont
			{Saitoh}}]{uchidaSurfaceacousticwavedrivenSpinPumping2011}%
		\BibitemOpen
		\bibfield  {author} {\bibinfo {author} {\bibfnamefont {K.-i.}\ \bibnamefont
				{Uchida}}, \bibinfo {author} {\bibfnamefont {T.}~\bibnamefont {An}}, \bibinfo
			{author} {\bibfnamefont {Y.}~\bibnamefont {Kajiwara}}, \bibinfo {author}
			{\bibfnamefont {M.}~\bibnamefont {Toda}},\ and\ \bibinfo {author}
			{\bibfnamefont {E.}~\bibnamefont {Saitoh}},\ }\bibfield  {title} {\bibinfo
			{title} {Surface-acoustic-wave-driven spin pumping in
				{{$\mathrm{Y_3Fe_5O_{12}}$/$\mathrm{Pt}$}} hybrid structure},\ }\href
		{https://doi.org/10.1063/1.3662032} {\bibfield  {journal} {\bibinfo
				{journal} {Applied Physics Letters}\ }\textbf {\bibinfo {volume} {99}},\
			\bibinfo {pages} {212501} (\bibinfo {year} {2011}{\natexlab{b}})}\BibitemShut
		{NoStop}%
		\bibitem [{\citenamefont {Sano}\ \emph {et~al.}(2024)\citenamefont {Sano},
			\citenamefont {Ominato},\ and\ \citenamefont
			{Matsuo}}]{sanoAcoustomagnonicSpinHall2024}%
		\BibitemOpen
		\bibfield  {author} {\bibinfo {author} {\bibfnamefont {R.}~\bibnamefont
				{Sano}}, \bibinfo {author} {\bibfnamefont {Y.}~\bibnamefont {Ominato}},\ and\
			\bibinfo {author} {\bibfnamefont {M.}~\bibnamefont {Matsuo}},\ }\bibfield
		{title} {\bibinfo {title} {Acoustomagnonic {{Spin Hall Effect}} in
				{{Honeycomb Antiferromagnets}}},\ }\href
		{https://doi.org/10.1103/PhysRevLett.132.236302} {\bibfield  {journal}
			{\bibinfo  {journal} {Physical Review Letters}\ }\textbf {\bibinfo {volume}
				{132}},\ \bibinfo {pages} {236302} (\bibinfo {year} {2024})}\BibitemShut
		{NoStop}%
		\bibitem [{\citenamefont {{\v S}mejkal}\ \emph {et~al.}(2023)\citenamefont {{\v
					S}mejkal}, \citenamefont {Marmodoro}, \citenamefont {Ahn}, \citenamefont
			{{Gonz{\'a}lez-Hern{\'a}ndez}}, \citenamefont {Turek}, \citenamefont
			{Mankovsky}, \citenamefont {Ebert}, \citenamefont {D'Souza}, \citenamefont
			{{\v S}ipr}, \citenamefont {Sinova},\ and\ \citenamefont
			{Jungwirth}}]{smejkalChiralMagnonsAltermagnetic2023}%
		\BibitemOpen
		\bibfield  {author} {\bibinfo {author} {\bibfnamefont {L.}~\bibnamefont {{\v
						S}mejkal}}, \bibinfo {author} {\bibfnamefont {A.}~\bibnamefont {Marmodoro}},
			\bibinfo {author} {\bibfnamefont {K.-H.}\ \bibnamefont {Ahn}}, \bibinfo
			{author} {\bibfnamefont {R.}~\bibnamefont {{Gonz{\'a}lez-Hern{\'a}ndez}}},
			\bibinfo {author} {\bibfnamefont {I.}~\bibnamefont {Turek}}, \bibinfo
			{author} {\bibfnamefont {S.}~\bibnamefont {Mankovsky}}, \bibinfo {author}
			{\bibfnamefont {H.}~\bibnamefont {Ebert}}, \bibinfo {author} {\bibfnamefont
				{S.~W.}\ \bibnamefont {D'Souza}}, \bibinfo {author} {\bibfnamefont
				{O.}~\bibnamefont {{\v S}ipr}}, \bibinfo {author} {\bibfnamefont
				{J.}~\bibnamefont {Sinova}},\ and\ \bibinfo {author} {\bibfnamefont
				{T.}~\bibnamefont {Jungwirth}},\ }\bibfield  {title} {\bibinfo {title}
			{Chiral {{Magnons}} in {{Altermagnetic}} {{${\mathrm{RuO}}_{2}$}}},\ }\href
		{https://doi.org/10.1103/PhysRevLett.131.256703} {\bibfield  {journal}
			{\bibinfo  {journal} {Physical Review Letters}\ }\textbf {\bibinfo {volume}
				{131}},\ \bibinfo {pages} {256703} (\bibinfo {year} {2023})}\BibitemShut
		{NoStop}%
		\bibitem [{\citenamefont {Liao}\ \emph {et~al.}(2024)\citenamefont {Liao},
			\citenamefont {Wang}, \citenamefont {Tien}, \citenamefont {Huang},\ and\
			\citenamefont {Qu}}]{liaoSeparationInverseAltermagnetic2024}%
		\BibitemOpen
		\bibfield  {author} {\bibinfo {author} {\bibfnamefont {C.-T.}\ \bibnamefont
				{Liao}}, \bibinfo {author} {\bibfnamefont {Y.-C.}\ \bibnamefont {Wang}},
			\bibinfo {author} {\bibfnamefont {Y.-C.}\ \bibnamefont {Tien}}, \bibinfo
			{author} {\bibfnamefont {S.-Y.}\ \bibnamefont {Huang}},\ and\ \bibinfo
			{author} {\bibfnamefont {D.}~\bibnamefont {Qu}},\ }\bibfield  {title}
		{\bibinfo {title} {Separation of {{Inverse Altermagnetic Spin-Splitting
						Effect}} from {{Inverse Spin Hall Effect}} in {{${\mathrm{RuO}}_{2}$}}},\
		}\href {https://doi.org/10.1103/PhysRevLett.133.056701} {\bibfield  {journal}
			{\bibinfo  {journal} {Physical Review Letters}\ }\textbf {\bibinfo {volume}
				{133}},\ \bibinfo {pages} {056701} (\bibinfo {year} {2024})}\BibitemShut
		{NoStop}%
		\bibitem [{\citenamefont {Kawada}\ \emph {et~al.}(2021)\citenamefont {Kawada},
			\citenamefont {Kawaguchi}, \citenamefont {Funato}, \citenamefont {Kohno},\
			and\ \citenamefont {Hayashi}}]{kawadaAcousticSpinHall2021}%
		\BibitemOpen
		\bibfield  {author} {\bibinfo {author} {\bibfnamefont {T.}~\bibnamefont
				{Kawada}}, \bibinfo {author} {\bibfnamefont {M.}~\bibnamefont {Kawaguchi}},
			\bibinfo {author} {\bibfnamefont {T.}~\bibnamefont {Funato}}, \bibinfo
			{author} {\bibfnamefont {H.}~\bibnamefont {Kohno}},\ and\ \bibinfo {author}
			{\bibfnamefont {M.}~\bibnamefont {Hayashi}},\ }\bibfield  {title} {\bibinfo
			{title} {Acoustic spin {{Hall}} effect in strong spin-orbit metals},\ }\href
		{https://doi.org/10.1126/sciadv.abd9697} {\bibfield  {journal} {\bibinfo
				{journal} {Science Advances}\ }\textbf {\bibinfo {volume} {7}},\ \bibinfo
			{pages} {eabd9697} (\bibinfo {year} {2021})}\BibitemShut {NoStop}%
		\bibitem [{\citenamefont {Bleustein}(1968)}]{bleusteinNewSurfaceWave1968}%
		\BibitemOpen
		\bibfield  {author} {\bibinfo {author} {\bibfnamefont {J.~L.}\ \bibnamefont
				{Bleustein}},\ }\bibfield  {title} {\bibinfo {title} {A new surface wave in
				piezoelectric materials},\ }\href {https://doi.org/10.1063/1.1652495}
		{\bibfield  {journal} {\bibinfo  {journal} {Applied Physics Letters}\
			}\textbf {\bibinfo {volume} {13}},\ \bibinfo {pages} {412} (\bibinfo {year}
			{1968})}\BibitemShut {NoStop}%
		\bibitem [{\citenamefont
			{Gulyaev}(1969)}]{gulyaevElectroacousticSurfaceWaves1969}%
		\BibitemOpen
		\bibfield  {author} {\bibinfo {author} {\bibfnamefont {{\relax Yu}.~V.}\
				\bibnamefont {Gulyaev}},\ }\bibfield  {title} {\bibinfo {title}
			{Electroacoustic surface waves in solids},\ }\href
		{http://jetpletters.ru/ps/0/article_25273.shtml} {\bibfield  {journal}
			{\bibinfo  {journal} {JETP Letters}\ }\textbf {\bibinfo {volume} {9}},\
			\bibinfo {pages} {37} (\bibinfo {year} {1969})}\BibitemShut {NoStop}%
		\bibitem [{Note1()}]{Note1}%
		\BibitemOpen
		\bibinfo {note} {See Supplemental Material at [URL will be inserted by
			publisher] for details on the derivation of the spin current in the metallic
			and insulating altermagnet, the detection of the transverse spin current, the
			surface acoustic wave, the $z$-component of the electric field, thickness
			effects and a possible experimental setup. Includes Refs.~\cite
			{andoElectronicPropertiesTwodimensional1982,kittelQuantumTheorySolids1987,ashcroftSolidStatePhysics1976,ruckriegelAngularMomentumConservation2020,troncosoSpinTransportThick2020,rezendeDiffusiveMagnonicSpin2016,yamanouchiPropagationAmplificationRayleigh1972,weisLithiumNiobateSummary1985,kingViscosityTensorApproach1969,fallMeasurementSurfaceAcoustic2025}.}\BibitemShut
		{Stop}%
		\bibitem [{\citenamefont {Dreher}\ \emph {et~al.}(2012)\citenamefont {Dreher},
			\citenamefont {Weiler}, \citenamefont {Pernpeintner}, \citenamefont {Huebl},
			\citenamefont {Gross}, \citenamefont {Brandt},\ and\ \citenamefont
			{Goennenwein}}]{dreherSurfaceAcousticWave2012}%
		\BibitemOpen
		\bibfield  {author} {\bibinfo {author} {\bibfnamefont {L.}~\bibnamefont
				{Dreher}}, \bibinfo {author} {\bibfnamefont {M.}~\bibnamefont {Weiler}},
			\bibinfo {author} {\bibfnamefont {M.}~\bibnamefont {Pernpeintner}}, \bibinfo
			{author} {\bibfnamefont {H.}~\bibnamefont {Huebl}}, \bibinfo {author}
			{\bibfnamefont {R.}~\bibnamefont {Gross}}, \bibinfo {author} {\bibfnamefont
				{M.~S.}\ \bibnamefont {Brandt}},\ and\ \bibinfo {author} {\bibfnamefont
				{S.~T.~B.}\ \bibnamefont {Goennenwein}},\ }\bibfield  {title} {\bibinfo
			{title} {Surface acoustic wave driven ferromagnetic resonance in nickel thin
				films: {{Theory}} and experiment},\ }\href
		{https://doi.org/10.1103/PhysRevB.86.134415} {\bibfield  {journal} {\bibinfo
				{journal} {Physical Review B}\ }\textbf {\bibinfo {volume} {86}},\ \bibinfo
			{pages} {134415} (\bibinfo {year} {2012})}\BibitemShut {NoStop}%
		\bibitem [{\citenamefont {Marder}(2010)}]{marderCondensedMatterPhysics2010}%
		\BibitemOpen
		\bibfield  {author} {\bibinfo {author} {\bibfnamefont {M.~P.}\ \bibnamefont
				{Marder}},\ }\href@noop {} {\emph {\bibinfo {title} {Condensed matter
					physics}}},\ \bibinfo {edition} {2nd}\ ed.\ (\bibinfo  {publisher} {Wiley},\
		\bibinfo {address} {Hoboken, N.J},\ \bibinfo {year} {2010})\BibitemShut
		{NoStop}%
		\bibitem [{\citenamefont {Lee}\ and\ \citenamefont
			{Spector}(1983)}]{leeDielectricFunctionQuasitwodimensional1983}%
		\BibitemOpen
		\bibfield  {author} {\bibinfo {author} {\bibfnamefont {J.}~\bibnamefont
				{Lee}}\ and\ \bibinfo {author} {\bibfnamefont {H.~N.}\ \bibnamefont
				{Spector}},\ }\bibfield  {title} {\bibinfo {title} {Dielectric function for a
				quasi-two-dimensional semiconducting system},\ }\href
		{https://doi.org/10.1063/1.332016} {\bibfield  {journal} {\bibinfo  {journal}
				{Journal of Applied Physics}\ }\textbf {\bibinfo {volume} {54}},\ \bibinfo
			{pages} {6989} (\bibinfo {year} {1983})}\BibitemShut {NoStop}%
		\bibitem [{\citenamefont {Gurevich}\ and\ \citenamefont
			{Melkov}(1996)}]{gurevichMagnetizationOscillationsWaves1996}%
		\BibitemOpen
		\bibfield  {author} {\bibinfo {author} {\bibfnamefont {A.~G.}\ \bibnamefont
				{Gurevich}}\ and\ \bibinfo {author} {\bibfnamefont {G.~A.}\ \bibnamefont
				{Melkov}},\ }\href@noop {} {\emph {\bibinfo {title} {Magnetization
					oscillations and waves}}}\ (\bibinfo  {publisher} {CRC Press},\ \bibinfo
		{address} {Boca Raton},\ \bibinfo {year} {1996})\BibitemShut {NoStop}%
		\bibitem [{\citenamefont {Weiler}\ \emph {et~al.}(2012)\citenamefont {Weiler},
			\citenamefont {Huebl}, \citenamefont {Goerg}, \citenamefont {Czeschka},
			\citenamefont {Gross},\ and\ \citenamefont
			{Goennenwein}}]{weilerSpinPumpingCoherent2012}%
		\BibitemOpen
		\bibfield  {author} {\bibinfo {author} {\bibfnamefont {M.}~\bibnamefont
				{Weiler}}, \bibinfo {author} {\bibfnamefont {H.}~\bibnamefont {Huebl}},
			\bibinfo {author} {\bibfnamefont {F.~S.}\ \bibnamefont {Goerg}}, \bibinfo
			{author} {\bibfnamefont {F.~D.}\ \bibnamefont {Czeschka}}, \bibinfo {author}
			{\bibfnamefont {R.}~\bibnamefont {Gross}},\ and\ \bibinfo {author}
			{\bibfnamefont {S.~T.~B.}\ \bibnamefont {Goennenwein}},\ }\bibfield  {title}
		{\bibinfo {title} {Spin {{Pumping}} with {{Coherent Elastic Waves}}},\ }\href
		{https://doi.org/10.1103/PhysRevLett.108.176601} {\bibfield  {journal}
			{\bibinfo  {journal} {Physical Review Letters}\ }\textbf {\bibinfo {volume}
				{108}},\ \bibinfo {pages} {176601} (\bibinfo {year} {2012})}\BibitemShut
		{NoStop}%
		\bibitem [{\citenamefont {Streib}\ \emph {et~al.}(2019)\citenamefont {Streib},
			\citenamefont {{Vidal-Silva}}, \citenamefont {Shen},\ and\ \citenamefont
			{Bauer}}]{streibMagnonphononInteractionsMagnetic2019}%
		\BibitemOpen
		\bibfield  {author} {\bibinfo {author} {\bibfnamefont {S.}~\bibnamefont
				{Streib}}, \bibinfo {author} {\bibfnamefont {N.}~\bibnamefont
				{{Vidal-Silva}}}, \bibinfo {author} {\bibfnamefont {K.}~\bibnamefont
				{Shen}},\ and\ \bibinfo {author} {\bibfnamefont {G.~E.~W.}\ \bibnamefont
				{Bauer}},\ }\bibfield  {title} {\bibinfo {title} {Magnon-phonon interactions
				in magnetic insulators},\ }\href {https://doi.org/10.1103/PhysRevB.99.184442}
		{\bibfield  {journal} {\bibinfo  {journal} {Physical Review B}\ }\textbf
			{\bibinfo {volume} {99}},\ \bibinfo {pages} {184442} (\bibinfo {year}
			{2019})}\BibitemShut {NoStop}%
		\bibitem [{\citenamefont {Cornelissen}\ \emph {et~al.}(2016)\citenamefont
			{Cornelissen}, \citenamefont {Peters}, \citenamefont {Bauer}, \citenamefont
			{Duine},\ and\ \citenamefont {{van
					Wees}}}]{cornelissenMagnonSpinTransport2016}%
		\BibitemOpen
		\bibfield  {author} {\bibinfo {author} {\bibfnamefont {L.~J.}\ \bibnamefont
				{Cornelissen}}, \bibinfo {author} {\bibfnamefont {K.~J.~H.}\ \bibnamefont
				{Peters}}, \bibinfo {author} {\bibfnamefont {G.~E.~W.}\ \bibnamefont
				{Bauer}}, \bibinfo {author} {\bibfnamefont {R.~A.}\ \bibnamefont {Duine}},\
			and\ \bibinfo {author} {\bibfnamefont {B.~J.}\ \bibnamefont {{van Wees}}},\
		}\bibfield  {title} {\bibinfo {title} {Magnon spin transport driven by the
				magnon chemical potential in a magnetic insulator},\ }\href
		{https://doi.org/10.1103/PhysRevB.94.014412} {\bibfield  {journal} {\bibinfo
				{journal} {Physical Review B}\ }\textbf {\bibinfo {volume} {94}},\ \bibinfo
			{pages} {014412} (\bibinfo {year} {2016})}\BibitemShut {NoStop}%
		\bibitem [{\citenamefont {Zhou}\ \emph {et~al.}(2024)\citenamefont {Zhou},
			\citenamefont {Feng}, \citenamefont {Zhang}, \citenamefont {{\v S}mejkal},
			\citenamefont {Sinova}, \citenamefont {Mokrousov},\ and\ \citenamefont
			{Yao}}]{zhouCrystalThermalTransport2024}%
		\BibitemOpen
		\bibfield  {author} {\bibinfo {author} {\bibfnamefont {X.}~\bibnamefont
				{Zhou}}, \bibinfo {author} {\bibfnamefont {W.}~\bibnamefont {Feng}}, \bibinfo
			{author} {\bibfnamefont {R.-W.}\ \bibnamefont {Zhang}}, \bibinfo {author}
			{\bibfnamefont {L.}~\bibnamefont {{\v S}mejkal}}, \bibinfo {author}
			{\bibfnamefont {J.}~\bibnamefont {Sinova}}, \bibinfo {author} {\bibfnamefont
				{Y.}~\bibnamefont {Mokrousov}},\ and\ \bibinfo {author} {\bibfnamefont
				{Y.}~\bibnamefont {Yao}},\ }\bibfield  {title} {\bibinfo {title} {Crystal
				{{Thermal Transport}} in {{Altermagnetic}} ${\mathrm{ruo}}_{2}$},\ }\href
		{https://doi.org/10.1103/PhysRevLett.132.056701} {\bibfield  {journal}
			{\bibinfo  {journal} {Physical Review Letters}\ }\textbf {\bibinfo {volume}
				{132}},\ \bibinfo {pages} {056701} (\bibinfo {year} {2024})}\BibitemShut
		{NoStop}%
		\bibitem [{\citenamefont {Bloch}(1966)}]{bloch10LawVolume1966}%
		\BibitemOpen
		\bibfield  {author} {\bibinfo {author} {\bibfnamefont {D.}~\bibnamefont
				{Bloch}},\ }\bibfield  {title} {\bibinfo {title} {The 10/3 law for the volume
				dependence of superexchange},\ }\href
		{https://doi.org/10.1016/0022-3697(66)90262-9} {\bibfield  {journal}
			{\bibinfo  {journal} {Journal of Physics and Chemistry of Solids}\ }\textbf
			{\bibinfo {volume} {27}},\ \bibinfo {pages} {881} (\bibinfo {year}
			{1966})}\BibitemShut {NoStop}%
		\bibitem [{\citenamefont {Ahn}\ \emph {et~al.}(2019)\citenamefont {Ahn},
			\citenamefont {Hariki}, \citenamefont {Lee},\ and\ \citenamefont {Kune{\v
					s}}}]{ahnAntiferromagnetism$mathrmRuO_2$$d$wave2019}%
		\BibitemOpen
		\bibfield  {author} {\bibinfo {author} {\bibfnamefont {K.-H.}\ \bibnamefont
				{Ahn}}, \bibinfo {author} {\bibfnamefont {A.}~\bibnamefont {Hariki}},
			\bibinfo {author} {\bibfnamefont {K.-W.}\ \bibnamefont {Lee}},\ and\ \bibinfo
			{author} {\bibfnamefont {J.}~\bibnamefont {Kune{\v s}}},\ }\bibfield  {title}
		{\bibinfo {title} {Antiferromagnetism in {{${\mathrm{RuO}}_{2}$}} as $d$-wave
				{{Pomeranchuk}} instability},\ }\href
		{https://doi.org/10.1103/PhysRevB.99.184432} {\bibfield  {journal} {\bibinfo
				{journal} {Physical Review B}\ }\textbf {\bibinfo {volume} {99}},\ \bibinfo
			{pages} {184432} (\bibinfo {year} {2019})}\BibitemShut {NoStop}%
		\bibitem [{\citenamefont {Tschirner}\ \emph {et~al.}(2023)\citenamefont
			{Tschirner}, \citenamefont {Ke{\ss}ler}, \citenamefont {Gonzalez~Betancourt},
			\citenamefont {Kotte}, \citenamefont {Kriegner}, \citenamefont {B{\"u}chner},
			\citenamefont {Dufouleur}, \citenamefont {Kamp}, \citenamefont {Jovic},
			\citenamefont {Smejkal}, \citenamefont {Sinova}, \citenamefont {Claessen},
			\citenamefont {Jungwirth}, \citenamefont {Moser}, \citenamefont {Reichlova},\
			and\ \citenamefont {Veyrat}}]{tschirnerSaturationAnomalousHall2023}%
		\BibitemOpen
		\bibfield  {author} {\bibinfo {author} {\bibfnamefont {T.}~\bibnamefont
				{Tschirner}}, \bibinfo {author} {\bibfnamefont {P.}~\bibnamefont
				{Ke{\ss}ler}}, \bibinfo {author} {\bibfnamefont {R.~D.}\ \bibnamefont
				{Gonzalez~Betancourt}}, \bibinfo {author} {\bibfnamefont {T.}~\bibnamefont
				{Kotte}}, \bibinfo {author} {\bibfnamefont {D.}~\bibnamefont {Kriegner}},
			\bibinfo {author} {\bibfnamefont {B.}~\bibnamefont {B{\"u}chner}}, \bibinfo
			{author} {\bibfnamefont {J.}~\bibnamefont {Dufouleur}}, \bibinfo {author}
			{\bibfnamefont {M.}~\bibnamefont {Kamp}}, \bibinfo {author} {\bibfnamefont
				{V.}~\bibnamefont {Jovic}}, \bibinfo {author} {\bibfnamefont
				{L.}~\bibnamefont {Smejkal}}, \bibinfo {author} {\bibfnamefont
				{J.}~\bibnamefont {Sinova}}, \bibinfo {author} {\bibfnamefont
				{R.}~\bibnamefont {Claessen}}, \bibinfo {author} {\bibfnamefont
				{T.}~\bibnamefont {Jungwirth}}, \bibinfo {author} {\bibfnamefont
				{S.}~\bibnamefont {Moser}}, \bibinfo {author} {\bibfnamefont
				{H.}~\bibnamefont {Reichlova}},\ and\ \bibinfo {author} {\bibfnamefont
				{L.}~\bibnamefont {Veyrat}},\ }\bibfield  {title} {\bibinfo {title}
			{Saturation of the anomalous {{Hall}} effect at high magnetic fields in
				altermagnetic {{${\mathrm{RuO}}_{2}$}}},\ }\href
		{https://doi.org/10.1063/5.0160335} {\bibfield  {journal} {\bibinfo
				{journal} {APL Materials}\ }\textbf {\bibinfo {volume} {11}},\ \bibinfo
			{pages} {101103} (\bibinfo {year} {2023})}\BibitemShut {NoStop}%
		\bibitem [{\citenamefont {Morano}\ \emph {et~al.}(2025)\citenamefont {Morano},
			\citenamefont {Maesen}, \citenamefont {Nikitin}, \citenamefont {Lass},
			\citenamefont {Mazzone},\ and\ \citenamefont
			{Zaharko}}]{moranoAbsenceAltermagneticMagnon2025}%
		\BibitemOpen
		\bibfield  {author} {\bibinfo {author} {\bibfnamefont {V.~C.}\ \bibnamefont
				{Morano}}, \bibinfo {author} {\bibfnamefont {Z.}~\bibnamefont {Maesen}},
			\bibinfo {author} {\bibfnamefont {S.~E.}\ \bibnamefont {Nikitin}}, \bibinfo
			{author} {\bibfnamefont {J.}~\bibnamefont {Lass}}, \bibinfo {author}
			{\bibfnamefont {D.~G.}\ \bibnamefont {Mazzone}},\ and\ \bibinfo {author}
			{\bibfnamefont {O.}~\bibnamefont {Zaharko}},\ }\bibfield  {title} {\bibinfo
			{title} {Absence of {{Altermagnetic Magnon Band Splitting}} in
				${\mathrm{mnf}}_{2}$},\ }\href
		{https://doi.org/10.1103/PhysRevLett.134.226702} {\bibfield  {journal}
			{\bibinfo  {journal} {Physical Review Letters}\ }\textbf {\bibinfo {volume}
				{134}},\ \bibinfo {pages} {226702} (\bibinfo {year} {2025})}\BibitemShut
		{NoStop}%
		\bibitem [{\citenamefont {Lebrun}\ \emph {et~al.}(2018)\citenamefont {Lebrun},
			\citenamefont {Ross}, \citenamefont {Bender}, \citenamefont {Qaiumzadeh},
			\citenamefont {Baldrati}, \citenamefont {Cramer}, \citenamefont {Brataas},
			\citenamefont {Duine},\ and\ \citenamefont
			{Kl{\"a}ui}}]{lebrunTunableLongdistanceSpin2018}%
		\BibitemOpen
		\bibfield  {author} {\bibinfo {author} {\bibfnamefont {R.}~\bibnamefont
				{Lebrun}}, \bibinfo {author} {\bibfnamefont {A.}~\bibnamefont {Ross}},
			\bibinfo {author} {\bibfnamefont {S.~A.}\ \bibnamefont {Bender}}, \bibinfo
			{author} {\bibfnamefont {A.}~\bibnamefont {Qaiumzadeh}}, \bibinfo {author}
			{\bibfnamefont {L.}~\bibnamefont {Baldrati}}, \bibinfo {author}
			{\bibfnamefont {J.}~\bibnamefont {Cramer}}, \bibinfo {author} {\bibfnamefont
				{A.}~\bibnamefont {Brataas}}, \bibinfo {author} {\bibfnamefont {R.~A.}\
				\bibnamefont {Duine}},\ and\ \bibinfo {author} {\bibfnamefont
				{M.}~\bibnamefont {Kl{\"a}ui}},\ }\bibfield  {title} {\bibinfo {title}
			{Tunable long-distance spin transport in a crystalline antiferromagnetic iron
				oxide},\ }\href {https://doi.org/10.1038/s41586-018-0490-7} {\bibfield
			{journal} {\bibinfo  {journal} {Nature}\ }\textbf {\bibinfo {volume} {561}},\
			\bibinfo {pages} {222} (\bibinfo {year} {2018})}\BibitemShut {NoStop}%
		\bibitem [{\citenamefont {Samara}\ and\ \citenamefont
			{Giardini}(1969)}]{samaraEffectPressureNeel1969}%
		\BibitemOpen
		\bibfield  {author} {\bibinfo {author} {\bibfnamefont {G.~A.}\ \bibnamefont
				{Samara}}\ and\ \bibinfo {author} {\bibfnamefont {A.~A.}\ \bibnamefont
				{Giardini}},\ }\bibfield  {title} {\bibinfo {title} {Effect of {{Pressure}}
				on the {{N\'eel Temperature}} of {{Magnetite}}},\ }\href
		{https://doi.org/10.1103/PhysRev.186.577} {\bibfield  {journal} {\bibinfo
				{journal} {Physical Review}\ }\textbf {\bibinfo {volume} {186}},\ \bibinfo
			{pages} {577} (\bibinfo {year} {1969})}\BibitemShut {NoStop}%
		\bibitem [{\citenamefont {Wang}\ \emph {et~al.}(2014)\citenamefont {Wang},
			\citenamefont {Du}, \citenamefont {Pu}, \citenamefont {Adur}, \citenamefont
			{Hammel},\ and\ \citenamefont {Yang}}]{wangScalingSpinHall2014}%
		\BibitemOpen
		\bibfield  {author} {\bibinfo {author} {\bibfnamefont {H.~L.}\ \bibnamefont
				{Wang}}, \bibinfo {author} {\bibfnamefont {C.~H.}\ \bibnamefont {Du}},
			\bibinfo {author} {\bibfnamefont {Y.}~\bibnamefont {Pu}}, \bibinfo {author}
			{\bibfnamefont {R.}~\bibnamefont {Adur}}, \bibinfo {author} {\bibfnamefont
				{P.~C.}\ \bibnamefont {Hammel}},\ and\ \bibinfo {author} {\bibfnamefont
				{F.~Y.}\ \bibnamefont {Yang}},\ }\bibfield  {title} {\bibinfo {title}
			{Scaling of {{Spin Hall Angle}} in 3d, 4d, and 5d {{Metals}} from
				{{$\mathrm{Y_3Fe_5O_{12}}$/Metal}} {{Spin Pumping}}},\ }\href
		{https://doi.org/10.1103/PhysRevLett.112.197201} {\bibfield  {journal}
			{\bibinfo  {journal} {Physical Review Letters}\ }\textbf {\bibinfo {volume}
				{112}},\ \bibinfo {pages} {197201} (\bibinfo {year} {2014})}\BibitemShut
		{NoStop}%
		\bibitem [{\citenamefont {Cui}\ \emph {et~al.}(2023)\citenamefont {Cui},
			\citenamefont {Zeng}, \citenamefont {Cui}, \citenamefont {Yu},\ and\
			\citenamefont {Yang}}]{cuiEfficientSpinSeebeck2023}%
		\BibitemOpen
		\bibfield  {author} {\bibinfo {author} {\bibfnamefont {Q.}~\bibnamefont
				{Cui}}, \bibinfo {author} {\bibfnamefont {B.}~\bibnamefont {Zeng}}, \bibinfo
			{author} {\bibfnamefont {P.}~\bibnamefont {Cui}}, \bibinfo {author}
			{\bibfnamefont {T.}~\bibnamefont {Yu}},\ and\ \bibinfo {author}
			{\bibfnamefont {H.}~\bibnamefont {Yang}},\ }\bibfield  {title} {\bibinfo
			{title} {Efficient spin {{Seebeck}} and spin {{Nernst}} effects of magnons in
				altermagnets},\ }\href {https://doi.org/10.1103/PhysRevB.108.L180401}
		{\bibfield  {journal} {\bibinfo  {journal} {Physical Review B}\ }\textbf
			{\bibinfo {volume} {108}},\ \bibinfo {pages} {L180401} (\bibinfo {year}
			{2023})}\BibitemShut {NoStop}%
		\bibitem [{\citenamefont {Yi}\ \emph {et~al.}(2025)\citenamefont {Yi},
			\citenamefont {Mao}, \citenamefont {Lu},\ and\ \citenamefont
			{Sun}}]{yiSpinSplittingNernst2025}%
		\BibitemOpen
		\bibfield  {author} {\bibinfo {author} {\bibfnamefont {X.-J.}\ \bibnamefont
				{Yi}}, \bibinfo {author} {\bibfnamefont {Y.}~\bibnamefont {Mao}}, \bibinfo
			{author} {\bibfnamefont {X.}~\bibnamefont {Lu}},\ and\ \bibinfo {author}
			{\bibfnamefont {Q.-F.}\ \bibnamefont {Sun}},\ }\bibfield  {title} {\bibinfo
			{title} {Spin splitting {{Nernst}} effect in altermagnets},\ }\href
		{https://doi.org/10.1103/PhysRevB.111.035423} {\bibfield  {journal} {\bibinfo
				{journal} {Physical Review B}\ }\textbf {\bibinfo {volume} {111}},\ \bibinfo
			{pages} {035423} (\bibinfo {year} {2025})}\BibitemShut {NoStop}%
		\bibitem [{\citenamefont {Amin}\ \emph {et~al.}(2024)\citenamefont {Amin},
			\citenamefont {Dal~Din}, \citenamefont {Golias}, \citenamefont {Niu},
			\citenamefont {Zakharov}, \citenamefont {Fromage}, \citenamefont {Fields},
			\citenamefont {Heywood}, \citenamefont {Cousins}, \citenamefont
			{Maccherozzi}, \citenamefont {Krempask{\'y}}, \citenamefont {Dil},
			\citenamefont {Kriegner}, \citenamefont {Kiraly}, \citenamefont {Campion},
			\citenamefont {Rushforth}, \citenamefont {Edmonds}, \citenamefont {Dhesi},
			\citenamefont {{\v S}mejkal}, \citenamefont {Jungwirth},\ and\ \citenamefont
			{Wadley}}]{aminNanoscaleImagingControl2024}%
		\BibitemOpen
		\bibfield  {author} {\bibinfo {author} {\bibfnamefont {O.~J.}\ \bibnamefont
				{Amin}}, \bibinfo {author} {\bibfnamefont {A.}~\bibnamefont {Dal~Din}},
			\bibinfo {author} {\bibfnamefont {E.}~\bibnamefont {Golias}}, \bibinfo
			{author} {\bibfnamefont {Y.}~\bibnamefont {Niu}}, \bibinfo {author}
			{\bibfnamefont {A.}~\bibnamefont {Zakharov}}, \bibinfo {author}
			{\bibfnamefont {S.~C.}\ \bibnamefont {Fromage}}, \bibinfo {author}
			{\bibfnamefont {C.~J.~B.}\ \bibnamefont {Fields}}, \bibinfo {author}
			{\bibfnamefont {S.~L.}\ \bibnamefont {Heywood}}, \bibinfo {author}
			{\bibfnamefont {R.~B.}\ \bibnamefont {Cousins}}, \bibinfo {author}
			{\bibfnamefont {F.}~\bibnamefont {Maccherozzi}}, \bibinfo {author}
			{\bibfnamefont {J.}~\bibnamefont {Krempask{\'y}}}, \bibinfo {author}
			{\bibfnamefont {J.~H.}\ \bibnamefont {Dil}}, \bibinfo {author} {\bibfnamefont
				{D.}~\bibnamefont {Kriegner}}, \bibinfo {author} {\bibfnamefont
				{B.}~\bibnamefont {Kiraly}}, \bibinfo {author} {\bibfnamefont {R.~P.}\
				\bibnamefont {Campion}}, \bibinfo {author} {\bibfnamefont {A.~W.}\
				\bibnamefont {Rushforth}}, \bibinfo {author} {\bibfnamefont {K.~W.}\
				\bibnamefont {Edmonds}}, \bibinfo {author} {\bibfnamefont {S.~S.}\
				\bibnamefont {Dhesi}}, \bibinfo {author} {\bibfnamefont {L.}~\bibnamefont
				{{\v S}mejkal}}, \bibinfo {author} {\bibfnamefont {T.}~\bibnamefont
				{Jungwirth}},\ and\ \bibinfo {author} {\bibfnamefont {P.}~\bibnamefont
				{Wadley}},\ }\bibfield  {title} {\bibinfo {title} {Nanoscale imaging and
				control of altermagnetism in {{MnTe}}},\ }\href
		{https://doi.org/10.1038/s41586-024-08234-x} {\bibfield  {journal} {\bibinfo
				{journal} {Nature}\ }\textbf {\bibinfo {volume} {636}},\ \bibinfo {pages}
			{348} (\bibinfo {year} {2024})}\BibitemShut {NoStop}%
		\bibitem [{\citenamefont {Jeong}\ \emph {et~al.}(2025)\citenamefont {Jeong},
			\citenamefont {Lee}, \citenamefont {Lin}, \citenamefont {Yang}, \citenamefont
			{Choi}, \citenamefont {Oh}, \citenamefont {Song}, \citenamefont {wook Lee},
			\citenamefont {Nair}, \citenamefont {Choudhary}, \citenamefont {Parikh},
			\citenamefont {Park}, \citenamefont {Choi}, \citenamefont {Lee},
			\citenamefont {LeBeau}, \citenamefont {Low},\ and\ \citenamefont
			{Jalan}}]{jeongMetallicityAnomalousHall2025}%
		\BibitemOpen
		\bibfield  {author} {\bibinfo {author} {\bibfnamefont {S.~G.}\ \bibnamefont
				{Jeong}}, \bibinfo {author} {\bibfnamefont {S.}~\bibnamefont {Lee}}, \bibinfo
			{author} {\bibfnamefont {B.}~\bibnamefont {Lin}}, \bibinfo {author}
			{\bibfnamefont {Z.}~\bibnamefont {Yang}}, \bibinfo {author} {\bibfnamefont
				{I.~H.}\ \bibnamefont {Choi}}, \bibinfo {author} {\bibfnamefont {J.~Y.}\
				\bibnamefont {Oh}}, \bibinfo {author} {\bibfnamefont {S.}~\bibnamefont
				{Song}}, \bibinfo {author} {\bibfnamefont {S.}~\bibnamefont {wook Lee}},
			\bibinfo {author} {\bibfnamefont {S.}~\bibnamefont {Nair}}, \bibinfo {author}
			{\bibfnamefont {R.}~\bibnamefont {Choudhary}}, \bibinfo {author}
			{\bibfnamefont {J.}~\bibnamefont {Parikh}}, \bibinfo {author} {\bibfnamefont
				{S.}~\bibnamefont {Park}}, \bibinfo {author} {\bibfnamefont {W.~S.}\
				\bibnamefont {Choi}}, \bibinfo {author} {\bibfnamefont {J.~S.}\ \bibnamefont
				{Lee}}, \bibinfo {author} {\bibfnamefont {J.~M.}\ \bibnamefont {LeBeau}},
			\bibinfo {author} {\bibfnamefont {T.}~\bibnamefont {Low}},\ and\ \bibinfo
			{author} {\bibfnamefont {B.}~\bibnamefont {Jalan}},\ }\href
		{https://doi.org/10.48550/arXiv.2501.11204} {\bibinfo {title} {Metallicity
				and {{Anomalous Hall Effect}} in {{Epitaxially-Strained}},
				{{Atomically-thin}} {{$\mathrm{RuO_2}$}} {{Films}}}} (\bibinfo {year}
		{2025}),\ \Eprint {https://arxiv.org/abs/2501.11204} {arXiv:2501.11204
			[cond-mat]} \BibitemShut {NoStop}%
		\bibitem [{\citenamefont {Reichlova}\ \emph {et~al.}(2024)\citenamefont
			{Reichlova}, \citenamefont {Lopes~Seeger}, \citenamefont
			{{Gonz{\'a}lez-Hern{\'a}ndez}}, \citenamefont {Kounta}, \citenamefont
			{Schlitz}, \citenamefont {Kriegner}, \citenamefont {Ritzinger}, \citenamefont
			{Lammel}, \citenamefont {Leivisk{\"a}}, \citenamefont {Birk~Hellenes},
			\citenamefont {Olejn{\'i}k}, \citenamefont {Pet{\v r}i{\v c}ek},
			\citenamefont {Dole{\v z}al}, \citenamefont {Horak}, \citenamefont
			{Schmoranzerova}, \citenamefont {Badura}, \citenamefont {Bertaina},
			\citenamefont {Thomas}, \citenamefont {Baltz}, \citenamefont {Michez},
			\citenamefont {Sinova}, \citenamefont {Goennenwein}, \citenamefont
			{Jungwirth},\ and\ \citenamefont {{\v
					S}mejkal}}]{reichlovaObservationSpontaneousAnomalous2024}%
		\BibitemOpen
		\bibfield  {author} {\bibinfo {author} {\bibfnamefont {H.}~\bibnamefont
				{Reichlova}}, \bibinfo {author} {\bibfnamefont {R.}~\bibnamefont
				{Lopes~Seeger}}, \bibinfo {author} {\bibfnamefont {R.}~\bibnamefont
				{{Gonz{\'a}lez-Hern{\'a}ndez}}}, \bibinfo {author} {\bibfnamefont
				{I.}~\bibnamefont {Kounta}}, \bibinfo {author} {\bibfnamefont
				{R.}~\bibnamefont {Schlitz}}, \bibinfo {author} {\bibfnamefont
				{D.}~\bibnamefont {Kriegner}}, \bibinfo {author} {\bibfnamefont
				{P.}~\bibnamefont {Ritzinger}}, \bibinfo {author} {\bibfnamefont
				{M.}~\bibnamefont {Lammel}}, \bibinfo {author} {\bibfnamefont
				{M.}~\bibnamefont {Leivisk{\"a}}}, \bibinfo {author} {\bibfnamefont
				{A.}~\bibnamefont {Birk~Hellenes}}, \bibinfo {author} {\bibfnamefont
				{K.}~\bibnamefont {Olejn{\'i}k}}, \bibinfo {author} {\bibfnamefont
				{V.}~\bibnamefont {Pet{\v r}i{\v c}ek}}, \bibinfo {author} {\bibfnamefont
				{P.}~\bibnamefont {Dole{\v z}al}}, \bibinfo {author} {\bibfnamefont
				{L.}~\bibnamefont {Horak}}, \bibinfo {author} {\bibfnamefont
				{E.}~\bibnamefont {Schmoranzerova}}, \bibinfo {author} {\bibfnamefont
				{A.}~\bibnamefont {Badura}}, \bibinfo {author} {\bibfnamefont
				{S.}~\bibnamefont {Bertaina}}, \bibinfo {author} {\bibfnamefont
				{A.}~\bibnamefont {Thomas}}, \bibinfo {author} {\bibfnamefont
				{V.}~\bibnamefont {Baltz}}, \bibinfo {author} {\bibfnamefont
				{L.}~\bibnamefont {Michez}}, \bibinfo {author} {\bibfnamefont
				{J.}~\bibnamefont {Sinova}}, \bibinfo {author} {\bibfnamefont {S.~T.~B.}\
				\bibnamefont {Goennenwein}}, \bibinfo {author} {\bibfnamefont
				{T.}~\bibnamefont {Jungwirth}},\ and\ \bibinfo {author} {\bibfnamefont
				{L.}~\bibnamefont {{\v S}mejkal}},\ }\bibfield  {title} {\bibinfo {title}
			{Observation of a spontaneous anomalous {{Hall}} response in the
				{{$\mathrm{Mn_5Si_3}$}} $d$-wave altermagnet candidate},\ }\href
		{https://doi.org/10.1038/s41467-024-48493-w} {\bibfield  {journal} {\bibinfo
				{journal} {Nature Communications}\ }\textbf {\bibinfo {volume} {15}},\
			\bibinfo {pages} {4961} (\bibinfo {year} {2024})}\BibitemShut {NoStop}%
		\bibitem [{\citenamefont {Chakraborty}\ \emph {et~al.}(2024)\citenamefont
			{Chakraborty}, \citenamefont {Gonz{\'a}lez~Hern{\'a}ndez}, \citenamefont {{\v
					S}mejkal},\ and\ \citenamefont
			{Sinova}}]{chakrabortyStraininducedPhaseTransition2024}%
		\BibitemOpen
		\bibfield  {author} {\bibinfo {author} {\bibfnamefont {A.}~\bibnamefont
				{Chakraborty}}, \bibinfo {author} {\bibfnamefont {R.}~\bibnamefont
				{Gonz{\'a}lez~Hern{\'a}ndez}}, \bibinfo {author} {\bibfnamefont
				{L.}~\bibnamefont {{\v S}mejkal}},\ and\ \bibinfo {author} {\bibfnamefont
				{J.}~\bibnamefont {Sinova}},\ }\bibfield  {title} {\bibinfo {title}
			{Strain-induced phase transition from antiferromagnet to altermagnet},\
		}\href {https://doi.org/10.1103/PhysRevB.109.144421} {\bibfield  {journal}
			{\bibinfo  {journal} {Physical Review B}\ }\textbf {\bibinfo {volume}
				{109}},\ \bibinfo {pages} {144421} (\bibinfo {year} {2024})}\BibitemShut
		{NoStop}%
		\bibitem [{\citenamefont {Karetta}\ \emph {et~al.}(2025)\citenamefont
			{Karetta}, \citenamefont {Verbeek}, \citenamefont {{Jaeschke-Ubiergo}},
			\citenamefont {{\v S}mejkal},\ and\ \citenamefont
			{Sinova}}]{karettaStraincontrolledWaveTransition2025}%
		\BibitemOpen
		\bibfield  {author} {\bibinfo {author} {\bibfnamefont {B.}~\bibnamefont
				{Karetta}}, \bibinfo {author} {\bibfnamefont {X.~H.}\ \bibnamefont
				{Verbeek}}, \bibinfo {author} {\bibfnamefont {R.}~\bibnamefont
				{{Jaeschke-Ubiergo}}}, \bibinfo {author} {\bibfnamefont {L.}~\bibnamefont
				{{\v S}mejkal}},\ and\ \bibinfo {author} {\bibfnamefont {J.}~\bibnamefont
				{Sinova}},\ }\bibfield  {title} {\bibinfo {title} {Strain-controlled g - to d
				-wave transition in altermagnetic {{CrSb}}},\ }\href
		{https://doi.org/10.1103/pbbr-hwz4} {\bibfield  {journal} {\bibinfo
				{journal} {Physical Review B}\ }\textbf {\bibinfo {volume} {112}},\ \bibinfo
			{pages} {094454} (\bibinfo {year} {2025})}\BibitemShut {NoStop}%
		\bibitem [{\citenamefont
			{Belashchenko}(2025)}]{belashchenkoGiantStrainInducedSpin2025}%
		\BibitemOpen
		\bibfield  {author} {\bibinfo {author} {\bibfnamefont {K.~D.}\ \bibnamefont
				{Belashchenko}},\ }\bibfield  {title} {\bibinfo {title} {Giant
				{{Strain-Induced Spin Splitting Effect}} in {{MnTe}}, a g -{{Wave
						Altermagnetic Semiconductor}}},\ }\href
		{https://doi.org/10.1103/PhysRevLett.134.086701} {\bibfield  {journal}
			{\bibinfo  {journal} {Physical Review Letters}\ }\textbf {\bibinfo {volume}
				{134}},\ \bibinfo {pages} {086701} (\bibinfo {year} {2025})}\BibitemShut
		{NoStop}%
		\bibitem [{\citenamefont {Li}\ \emph {et~al.}(2025)\citenamefont {Li},
			\citenamefont {Hu}, \citenamefont {Zhang}, \citenamefont {Berntsen},
			\citenamefont {Scali}, \citenamefont {Phuyal}, \citenamefont {Lin},
			\citenamefont {Chen}, \citenamefont {Chang}, \citenamefont {Clark},
			\citenamefont {Kim}, \citenamefont {Osiecki}, \citenamefont {Polley},
			\citenamefont {Thiagarajan}, \citenamefont {Li}, \citenamefont {Xiang},\ and\
			\citenamefont {Tjernberg}}]{liExtremeStrainControlled2025}%
		\BibitemOpen
		\bibfield  {author} {\bibinfo {author} {\bibfnamefont {C.}~\bibnamefont
				{Li}}, \bibinfo {author} {\bibfnamefont {M.}~\bibnamefont {Hu}}, \bibinfo
			{author} {\bibfnamefont {J.}~\bibnamefont {Zhang}}, \bibinfo {author}
			{\bibfnamefont {M.~H.}\ \bibnamefont {Berntsen}}, \bibinfo {author}
			{\bibfnamefont {F.}~\bibnamefont {Scali}}, \bibinfo {author} {\bibfnamefont
				{D.}~\bibnamefont {Phuyal}}, \bibinfo {author} {\bibfnamefont
				{C.}~\bibnamefont {Lin}}, \bibinfo {author} {\bibfnamefont {W.}~\bibnamefont
				{Chen}}, \bibinfo {author} {\bibfnamefont {J.}~\bibnamefont {Chang}},
			\bibinfo {author} {\bibfnamefont {O.~J.}\ \bibnamefont {Clark}}, \bibinfo
			{author} {\bibfnamefont {T.~K.}\ \bibnamefont {Kim}}, \bibinfo {author}
			{\bibfnamefont {J.}~\bibnamefont {Osiecki}}, \bibinfo {author} {\bibfnamefont
				{C.}~\bibnamefont {Polley}}, \bibinfo {author} {\bibfnamefont
				{B.}~\bibnamefont {Thiagarajan}}, \bibinfo {author} {\bibfnamefont
				{Z.}~\bibnamefont {Li}}, \bibinfo {author} {\bibfnamefont {T.}~\bibnamefont
				{Xiang}},\ and\ \bibinfo {author} {\bibfnamefont {O.}~\bibnamefont
				{Tjernberg}},\ }\href {https://doi.org/10.48550/arXiv.2512.07683} {\bibinfo
			{title} {Extreme {{Strain Controlled Correlated Metal-Insulator Transition}}
				in the {{Altermagnet CrSb}}}} (\bibinfo {year} {2025}),\ \Eprint
		{https://arxiv.org/abs/2512.07683} {arXiv:2512.07683 [cond-mat]} \BibitemShut
		{NoStop}%
		\bibitem [{\citenamefont {Zhou}\ \emph {et~al.}(2025)\citenamefont {Zhou},
			\citenamefont {Cheng}, \citenamefont {Hu}, \citenamefont {Chu}, \citenamefont
			{Bai}, \citenamefont {Han}, \citenamefont {Liu}, \citenamefont {Pan},\ and\
			\citenamefont {Song}}]{zhouManipulationAltermagneticOrder2025}%
		\BibitemOpen
		\bibfield  {author} {\bibinfo {author} {\bibfnamefont {Z.}~\bibnamefont
				{Zhou}}, \bibinfo {author} {\bibfnamefont {X.}~\bibnamefont {Cheng}},
			\bibinfo {author} {\bibfnamefont {M.}~\bibnamefont {Hu}}, \bibinfo {author}
			{\bibfnamefont {R.}~\bibnamefont {Chu}}, \bibinfo {author} {\bibfnamefont
				{H.}~\bibnamefont {Bai}}, \bibinfo {author} {\bibfnamefont {L.}~\bibnamefont
				{Han}}, \bibinfo {author} {\bibfnamefont {J.}~\bibnamefont {Liu}}, \bibinfo
			{author} {\bibfnamefont {F.}~\bibnamefont {Pan}},\ and\ \bibinfo {author}
			{\bibfnamefont {C.}~\bibnamefont {Song}},\ }\bibfield  {title} {\bibinfo
			{title} {Manipulation of the altermagnetic order in {{CrSb}} via crystal
				symmetry},\ }\href {https://doi.org/10.1038/s41586-024-08436-3} {\bibfield
			{journal} {\bibinfo  {journal} {Nature}\ }\textbf {\bibinfo {volume} {638}},\
			\bibinfo {pages} {645} (\bibinfo {year} {2025})}\BibitemShut {NoStop}%
		\bibitem [{\citenamefont {Gunnink}\ \emph {et~al.}(2026)\citenamefont
			{Gunnink}, \citenamefont {Sinova},\ and\ \citenamefont
			{Mook}}]{gunninkSurfaceAcousticWave2026}%
		\BibitemOpen
		\bibfield  {author} {\bibinfo {author} {\bibfnamefont {P.}~\bibnamefont
				{Gunnink}}, \bibinfo {author} {\bibfnamefont {J.}~\bibnamefont {Sinova}},\
			and\ \bibinfo {author} {\bibfnamefont {A.}~\bibnamefont {Mook}},\ }\href
		{https://doi.org/10.5281/zenodo.18315543} {\bibinfo {title} {Surface acoustic
				wave driven acoustic spin splitter in d-wave altermagnetic thin films:
				code}},\ \bibinfo {howpublished} {Zenodo} (\bibinfo {year}
		{2026})\BibitemShut {NoStop}%
		\bibitem [{\citenamefont {Ando}\ \emph {et~al.}(1982)\citenamefont {Ando},
			\citenamefont {Fowler},\ and\ \citenamefont
			{Stern}}]{andoElectronicPropertiesTwodimensional1982}%
		\BibitemOpen
		\bibfield  {author} {\bibinfo {author} {\bibfnamefont {T.}~\bibnamefont
				{Ando}}, \bibinfo {author} {\bibfnamefont {A.~B.}\ \bibnamefont {Fowler}},\
			and\ \bibinfo {author} {\bibfnamefont {F.}~\bibnamefont {Stern}},\ }\bibfield
		{title} {\bibinfo {title} {Electronic properties of two-dimensional
				systems},\ }\href {https://doi.org/10.1103/RevModPhys.54.437} {\bibfield
			{journal} {\bibinfo  {journal} {Reviews of Modern Physics}\ }\textbf
			{\bibinfo {volume} {54}},\ \bibinfo {pages} {437} (\bibinfo {year}
			{1982})}\BibitemShut {NoStop}%
		\bibitem [{\citenamefont {Kittel}\ and\ \citenamefont
			{Fong}(1987)}]{kittelQuantumTheorySolids1987}%
		\BibitemOpen
		\bibfield  {author} {\bibinfo {author} {\bibfnamefont {C.}~\bibnamefont
				{Kittel}}\ and\ \bibinfo {author} {\bibfnamefont {C.~Y.}\ \bibnamefont
				{Fong}},\ }\href@noop {} {\emph {\bibinfo {title} {Quantum theory of
					solids}}},\ \bibinfo {edition} {2nd}\ ed.\ (\bibinfo  {publisher} {Wiley},\
		\bibinfo {address} {New York},\ \bibinfo {year} {1987})\BibitemShut {NoStop}%
		\bibitem [{\citenamefont {Ashcroft}\ and\ \citenamefont
			{Mermin}(1976)}]{ashcroftSolidStatePhysics1976}%
		\BibitemOpen
		\bibfield  {author} {\bibinfo {author} {\bibfnamefont {N.~W.}\ \bibnamefont
				{Ashcroft}}\ and\ \bibinfo {author} {\bibfnamefont {N.~D.}\ \bibnamefont
				{Mermin}},\ }\href@noop {} {\emph {\bibinfo {title} {Solid {{State
							Physics}}}}}\ (\bibinfo  {publisher} {Saunders College Publishing},\ \bibinfo
		{year} {1976})\BibitemShut {NoStop}%
		\bibitem [{\citenamefont {R{\"u}ckriegel}\ \emph {et~al.}(2020)\citenamefont
			{R{\"u}ckriegel}, \citenamefont {Streib}, \citenamefont {Bauer},\ and\
			\citenamefont {Duine}}]{ruckriegelAngularMomentumConservation2020}%
		\BibitemOpen
		\bibfield  {author} {\bibinfo {author} {\bibfnamefont {A.}~\bibnamefont
				{R{\"u}ckriegel}}, \bibinfo {author} {\bibfnamefont {S.}~\bibnamefont
				{Streib}}, \bibinfo {author} {\bibfnamefont {G.~E.~W.}\ \bibnamefont
				{Bauer}},\ and\ \bibinfo {author} {\bibfnamefont {R.~A.}\ \bibnamefont
				{Duine}},\ }\bibfield  {title} {\bibinfo {title} {Angular {{Momentum
						Conservation}} and {{Phonon Spin}} in {{Magnetic Insulators}}},\ }\href
		{https://doi.org/10.1103/PhysRevB.101.104402} {\bibfield  {journal} {\bibinfo
				{journal} {Physical Review B}\ }\textbf {\bibinfo {volume} {101}},\ \bibinfo
			{pages} {104402} (\bibinfo {year} {2020})}\BibitemShut {NoStop}%
		\bibitem [{\citenamefont {Troncoso}\ \emph {et~al.}(2020)\citenamefont
			{Troncoso}, \citenamefont {Bender}, \citenamefont {Brataas},\ and\
			\citenamefont {Duine}}]{troncosoSpinTransportThick2020}%
		\BibitemOpen
		\bibfield  {author} {\bibinfo {author} {\bibfnamefont {R.~E.}\ \bibnamefont
				{Troncoso}}, \bibinfo {author} {\bibfnamefont {S.~A.}\ \bibnamefont
				{Bender}}, \bibinfo {author} {\bibfnamefont {A.}~\bibnamefont {Brataas}},\
			and\ \bibinfo {author} {\bibfnamefont {R.~A.}\ \bibnamefont {Duine}},\
		}\bibfield  {title} {\bibinfo {title} {Spin transport in thick insulating
				antiferromagnetic films},\ }\href
		{https://doi.org/10.1103/PhysRevB.101.054404} {\bibfield  {journal} {\bibinfo
				{journal} {Physical Review B}\ }\textbf {\bibinfo {volume} {101}},\ \bibinfo
			{pages} {054404} (\bibinfo {year} {2020})}\BibitemShut {NoStop}%
		\bibitem [{\citenamefont {Rezende}\ \emph {et~al.}(2016)\citenamefont
			{Rezende}, \citenamefont {{Rodr{\'i}guez-Su{\'a}rez}},\ and\ \citenamefont
			{Azevedo}}]{rezendeDiffusiveMagnonicSpin2016}%
		\BibitemOpen
		\bibfield  {author} {\bibinfo {author} {\bibfnamefont {S.~M.}\ \bibnamefont
				{Rezende}}, \bibinfo {author} {\bibfnamefont {R.~L.}\ \bibnamefont
				{{Rodr{\'i}guez-Su{\'a}rez}}},\ and\ \bibinfo {author} {\bibfnamefont
				{A.}~\bibnamefont {Azevedo}},\ }\bibfield  {title} {\bibinfo {title}
			{Diffusive magnonic spin transport in antiferromagnetic insulators},\ }\href
		{https://doi.org/10.1103/PhysRevB.93.054412} {\bibfield  {journal} {\bibinfo
				{journal} {Physical Review B}\ }\textbf {\bibinfo {volume} {93}},\ \bibinfo
			{pages} {054412} (\bibinfo {year} {2016})}\BibitemShut {NoStop}%
		\bibitem [{\citenamefont {Yamanouchi}\ and\ \citenamefont
			{Shibayama}(1972)}]{yamanouchiPropagationAmplificationRayleigh1972}%
		\BibitemOpen
		\bibfield  {author} {\bibinfo {author} {\bibfnamefont {K.}~\bibnamefont
				{Yamanouchi}}\ and\ \bibinfo {author} {\bibfnamefont {K.}~\bibnamefont
				{Shibayama}},\ }\bibfield  {title} {\bibinfo {title} {Propagation and
				{{Amplification}} of {{Rayleigh Waves}} and {{Piezoelectric Leaky Surface
						Waves}} in {{${\mathrm{LiNbO_3}}$}}},\ }\href
		{https://doi.org/10.1063/1.1661294} {\bibfield  {journal} {\bibinfo
				{journal} {Journal of Applied Physics}\ }\textbf {\bibinfo {volume} {43}},\
			\bibinfo {pages} {856} (\bibinfo {year} {1972})}\BibitemShut {NoStop}%
		\bibitem [{\citenamefont {Weis}\ and\ \citenamefont
			{Gaylord}(1985)}]{weisLithiumNiobateSummary1985}%
		\BibitemOpen
		\bibfield  {author} {\bibinfo {author} {\bibfnamefont {R.~S.}\ \bibnamefont
				{Weis}}\ and\ \bibinfo {author} {\bibfnamefont {T.~K.}\ \bibnamefont
				{Gaylord}},\ }\bibfield  {title} {\bibinfo {title} {Lithium niobate:
				{{Summary}} of physical properties and crystal structure},\ }\href
		{https://doi.org/10.1007/BF00614817} {\bibfield  {journal} {\bibinfo
				{journal} {Applied Physics A}\ }\textbf {\bibinfo {volume} {37}},\ \bibinfo
			{pages} {191} (\bibinfo {year} {1985})}\BibitemShut {NoStop}%
		\bibitem [{\citenamefont {King}\ and\ \citenamefont
			{Sheard}(1969)}]{kingViscosityTensorApproach1969}%
		\BibitemOpen
		\bibfield  {author} {\bibinfo {author} {\bibfnamefont {P.~J.}\ \bibnamefont
				{King}}\ and\ \bibinfo {author} {\bibfnamefont {F.~W.}\ \bibnamefont
				{Sheard}},\ }\bibfield  {title} {\bibinfo {title} {Viscosity {{Tensor
						Approach}} to the {{Damping}} of {{Rayleigh Waves}}},\ }\href
		{https://doi.org/10.1063/1.1657373} {\bibfield  {journal} {\bibinfo
				{journal} {Journal of Applied Physics}\ }\textbf {\bibinfo {volume} {40}},\
			\bibinfo {pages} {5189} (\bibinfo {year} {1969})}\BibitemShut {NoStop}%
		\bibitem [{\citenamefont {Fall}\ \emph {et~al.}(2025)\citenamefont {Fall},
			\citenamefont {Duquennoy}, \citenamefont {Smagin}, \citenamefont
			{Oumekloul},\ and\ \citenamefont
			{Ouaftouh}}]{fallMeasurementSurfaceAcoustic2025}%
		\BibitemOpen
		\bibfield  {author} {\bibinfo {author} {\bibfnamefont {D.}~\bibnamefont
				{Fall}}, \bibinfo {author} {\bibfnamefont {M.}~\bibnamefont {Duquennoy}},
			\bibinfo {author} {\bibfnamefont {N.}~\bibnamefont {Smagin}}, \bibinfo
			{author} {\bibfnamefont {Z.}~\bibnamefont {Oumekloul}},\ and\ \bibinfo
			{author} {\bibfnamefont {M.}~\bibnamefont {Ouaftouh}},\ }\bibfield  {title}
		{\bibinfo {title} {Measurement of surface acoustic wave attenuation in
				aluminum using transverse acoustic field},\ }\href
		{https://doi.org/10.1063/5.0260423} {\bibfield  {journal} {\bibinfo
				{journal} {AIP Advances}\ }\textbf {\bibinfo {volume} {15}},\ \bibinfo
			{pages} {065002} (\bibinfo {year} {2025})}\BibitemShut {NoStop}%
	\end{thebibliography}
\end{document}